%% file: main.tex
\pdfoutput=1
\documentclass[runningheads]{llncs}

\usepackage[T1]{fontenc}

\usepackage{amsmath}
\usepackage{amssymb}

\usepackage{newtxtext}
\usepackage[varvw]{newtxmath}

\usepackage{graphicx}
\usepackage{float}
\usepackage{booktabs}
\usepackage{tabularx}
\usepackage{longtable}

\usepackage{algpseudocode}
\usepackage{xurl}
\usepackage[hidelinks,bookmarksnumbered]{hyperref}
\usepackage[capitalise,noabbrev]{cleveref}
\usepackage{acro}
\DeclareAcronym{amm}{short=AMM, long=automated market maker}
\DeclareAcronym{defi}{short=DeFi, long=decentralized finance}

\input{numbers}

\input{table6_rows}
\input{table_live_rows}

\crefname{appendix}{Appendix}{Appendices}
\Crefname{appendix}{Appendix}{Appendices}

\newcommand{\lltv}{\lambda}
\newcommand{\booked}{\hat{p}}
\newcommand{\pbar}{\bar{p}}
\newcommand{\liq}{\ell}
\newcommand{\dd}{\mathrm{d}}
\newcommand{\pos}[1]{\left[#1\right]^{+}}

\begin{document}

\title{When the Peg Breaks: Modeling and Measuring Stablecoin Depeg Losses across DeFi}

\author{Weilin Li \and Jiahua Xu}
\authorrunning{W. Li and J. Xu}
\institute{University College London}

\maketitle

\begin{abstract}

Stablecoins and pegged coins are integrated deep into DeFi: pledged as collateral in lending markets, paired on AMMs, reached through vaults that lend to those markets, and held in the reserves of other coins. When such a coin loses its peg, these connections carry the loss to parties who never held it, and make it larger. Prior studies of depegs trace the coin's price, its holders' flows and causes. We give, to our knowledge, the first systematic study of how DeFi integration spreads and enlarges the losses of a depeg: we model the places a coin reaches as a graph of DeFi protocols, and derive for each kind of connection who loses and how much. The largest term arises in lending markets that fix the coin's oracle price at par: anyone can buy the coin cheaply, borrow real assets against it and never repay, leaving the loss with the lenders (oracle arbitrage). Empirically, we studied the depeg events over the last three years. \nPcUnionSizedFamilies{} stablecoins and pegged tokens over \$10M market cap depegged. In \nCensusEvents{} depeg events, lending markets kept pricing the coin above the market, and these events caused a total loss of \nDedupTotal{}: \nDedupCreditPct{} fell on lenders who never held the coin, \nDedupFreeFloatPct{} on holders, \nDedupBorrowerPct{} on borrowers and \nDedupPoolsPct{} on AMM liquidity providers. Every term of the bound is public chain state, so a coin's worst case can be computed before its depeg. Among the \nLiveBoundRowsListed{} largest of the pegged coins listed in lending markets, a fall to half of par would result in a loss from \nLiveBoundRankListedLastAtFifty{} (\nLiveBoundRankListedLastCoin{}) to \nLiveBoundTopOneAtFifty{} (\nLiveBoundTopOneCoin{}), and would cost the coin's lenders up to \nLiveBoundLenderTopOneValue{} (\nLiveBoundLenderTopOneCoin{}).
\end{abstract}

\section{Introduction}\label{sec:intro}

A stablecoin, or more generally a pegged coin, promises a fixed value in dollars or in another asset such as ETH. However, many fail to keep the peg for various reasons: an unauthorized mint, a loss in the reserves, or a failed verification of the reserves. Between \nPcWindowFromMonth{} and \nPcWindowToMonth{}, \nPcUnionSizedFamilies{} pegged coins with more than \$10M market cap traded below their peg on chain, with a combined peak market cap of \nPcUnionPeakSizeUsd{}. A natural reading is that each coin's holders bore the losses, in proportion to its fall.

In fact, the loss travels further. Many of these coins are integrated deep into \ac{defi}: pledged as collateral in lending markets, paired with other coins on \acp{amm}, reached through vaults that lend to those markets, and held in the reserves of funds and other stablecoins. Each of these integrations ties other people's money to the coin so the loss of a depeg can reach people who never held it, as shown in the two costliest depegs of the window. Stream's xUSD in November 2025 cost \nStreamDedup{}, of which \nStreamDepositor{} fell on the depositors of lending markets that had taken xUSD as collateral. Kelp's rsETH in April 2026 cost \nRsethDedup{}, all of it on the depositors of Aave's pools, who had supplied WETH and wstETH rather than rsETH, while the holders lost nothing lasting. Who loses in a depeg, and how much, depends on where the coin was integrated and on how each of those places prices it.

These connections pass the loss on. A lending market prices the collateral it lends against through an oracle, and for pegged coins many markets fix the oracle price at par, or at the exchange rate of a staked version. The choice is deliberate: by assuming the peg, it needs no price feed, and market noise cannot liquidate healthy borrowers. Once the market price falls below the oracle price, however, anyone can buy the coin cheaply, pledge it, borrow real dollars against the oracle price, and never repay. We call this trade \emph{oracle arbitrage}. The depositors are left with collateral that sells only at the market price, so the loss exceeds the holders' loss and falls on people who never held the coin. Under an oracle that follows the market, the loss falls instead on borrowers, who are liquidated, and on depositors where liquidity is too thin to liquidate in time. An \ac{amm} built for coins that trade near par buys the falling coin at close to par until its other side runs out, and leaves its liquidity providers holding only the depegged coin. A vault that lends to a hit market passes its share of the loss to its own depositors, and a fund or stablecoin that holds the coin in its reserves loses in proportion to that share.

The damage cascades beyond the parties who funded the coin. A market emptied by oracle arbitrage sits at full utilization. Its depositors cannot withdraw, and its borrow rate climbs to its cap. The vaults that funded it pay withdrawals from their other markets. Those markets fill up too, and their borrowers pay the capped rate on unrelated collateral. Depositors then leave vaults that share only a manager under panic. A coin that holds the depegged coin in its reserves can lose its own peg, and the cascade starts again.

We ask how the loss of a depeg travels through \ac{defi}, and how much each party along the way loses. Prior work does not answer this. Studies of stablecoin failures explain why a peg breaks and who runs first~\cite{liu2023anatomy,ma2025runs,saengchote2021run}, not who lost across DeFi protocols. Studies of \ac{defi} lending measure liquidations and oracle manipulation under oracles that follow the market~\cite{qin2021liquidations,perez2021knife,eskandari2021sok}, whereas the losses above arise when a market fixes the price. A case study of the November 2025 cascade~\cite{wen2026contagion} finds \ac{defi} lending exposure negligible. In our broader scan the exposure is far larger (\cref{sec:related}).

To our knowledge, this paper is the first systematic study of how \ac{defi} integration spreads and enlarges the losses of a depeg. We model the places a coin reaches as a graph, derive each party's loss from it, bound the worst case from public chain state, and test both on three years of depegs. We make three contributions.
\vspace{-0.13cm}
\begin{itemize}
\item \textbf{A systematic model of how a depeg's loss travels through \ac{defi}}. We model the places a coin reaches as a graph and derive, for each kind of connection, who loses as the coin falls to a price $p$, from which price, and how much. The model also follows the cascade: an emptied market pins its depositors and its rate, the vaults that funded it carry the damage to their other markets and to borrowers with no position in the coin, and a shared manager or a reserve holding carries it further.
\item \textbf{A worst-case loss bound from public state before a depeg}. Every term of the bound is public chain state, so the worst case can be computed given a snapshot state. Each term is a parameter that someone sets, which tells how to lower the exposure. Run today, it gives the exposure that stands for the pegged coins. Among the \nLiveBoundRowsListed{} largest, a fall to half of par would result in a loss from \nLiveBoundRankListedLastAtFifty{} (\nLiveBoundRankListedLastCoin{}) to \nLiveBoundTopOneAtFifty{} (\nLiveBoundTopOneCoin{}), and would cost the coin's lenders up to \nLiveBoundLenderTopOneValue{} (wstETH).
\item \textbf{An empirical study of three years of depegs}. Scanning every stablecoin and pegged token over \$10M market cap, we find \nPcUnionSizedFamilies{} that depegged between \nPcWindowFromMonth{} and \nPcWindowToMonth{}. In \nCensusEvents{} events, found by a census of \nCensusProtocols{} lending protocols, lending markets priced the coin above the market, and we apply the model to all of them. We reconstruct five events position by position, and attribute the loss party by party: of \nDedupTotalFive{}, \nDepositorSideFive{} fell on depositors who never held the coin.
\end{itemize}

\section{Background and Related Work}\label{sec:background}

\subsection{Stablecoins and their staked forms}\label{sec:bg-coins}

Stablecoins hold their peg in several ways~\cite{moin2020sok,ling2025lego}. A common one is an issuer that holds reserves and redeems at par for whitelisted users. Others trade on secondary markets, where arbitrage keeps the price near par~\cite{lyons2020keeps,ma2025runs,hui2025peg}, so the redemption price and market price can differ~\cite{eichengreen2025devaluation}. This paper does not depend on the design: the model needs only the market price and the price a protocol books, and applies to any pegged coin, whether its peg rests on reserves, on crypto collateral~\cite{klages2022stability,kozhan2021dai} or on an algorithm, including tokens pegged to tokens other than USD (e.g., stETH pegged to ETH)~\cite{xiong2024leverage}. Over half of active stablecoins carry a yield mechanism~\cite{ling2025lego}, often paid through a staked version: the coin, $\mathrm{USD}_A$, is deposited into an ERC-4626 contract~\cite{erc4626}, $\mathrm{sUSD}_A$, whose exchange rate rises as yield accrues. Mostly, the staked version is the form pledged as collateral.

\subsection{\acs{defi} systems}\label{sec:bg-lending}

\noindent\textit{Lending.} A lending market lends deposits of a loan asset against collateral~\cite{bartoletti2021lending}. A position whose debt exceeds the liquidation threshold can be liquidated: anyone repays part of the debt and takes collateral worth more by the liquidation penalty~\cite{qin2021liquidations}. A position whose debt exceeds its collateral's value is bad debt, which the depositors bear. A supply cap limits the collateral a market accepts. The borrow rate rises with utilization, the share of deposits lent out. At full utilization no depositor can withdraw, and some rate models keep raising it up to a cap~\cite{morphoirm}. Isolated markets (Morpho, Euler, Silo~\cite{morphoblue,eulerv2}) confine bad debt to one market's depositors, and shared pools (Aave, Spark~\cite{aavev3}) spread it over every depositor. A looper pledges $\mathrm{sUSD}_A$, borrows, buys more and repeats~\cite{xiong2024leverage}.

\noindent\textit{Oracles.} Each market designates an oracle contract that reports the collateral price it uses~\cite{eskandari2021sok}: an \ac{amm} or off-chain feed, an exchange rate, or a constant (\cref{sec:system}).

\noindent\textit{Automated market makers.} An \ac{amm} holds two coins and quotes a price from its balances~\cite{xu2023sokamm,angeris2020cfmm}. Stablecoin \acp{amm} quote near par while balanced~\cite{curve,uniswapv3,velodrome}, and when one coin falls they absorb it until the other side runs out, at the providers' expense~\cite{heimbach2021lp,milionis2022lvr}.

\noindent\textit{Vaults.} Most depositors use a vault, an ERC-4626 contract whose manager (curator) picks its markets and caps each. One manager often runs several vaults~\cite{zbandut2025curation}, and an allocator may move idle deposits between those markets~\cite{allocator}.

\subsection{Related work}\label{sec:related}

Empirical studies of stablecoin failure trace prices and flows: who ran first in the Terra collapse~\cite{liu2023anatomy}, redemption arbitrage and runs~\cite{ma2025runs}, spillovers across coins~\cite{gadzinski2024break,wu2026tail}, \ac{amm} liquidity and transfer networks in the March 2023 USDC depeg~\cite{hernandez2024noquestions,sankaewtong2026tracing}, and an early on-chain run~\cite{saengchote2021run}. They measure who ran and at what price, not who in \ac{defi} protocols bore the loss. Studies of \ac{defi} lending measure liquidations and their incentives~\cite{qin2021liquidations,perez2021knife}, rates and withdrawals at full utilization~\cite{heimbach2023merge}, bad debt manufactured through one position~\cite{heimbach2023squeeze}, leverage loops~\cite{wang2022speculative,xiong2024leverage}, and the price feeds markets read~\cite{eskandari2021sok}. Fixed or hard-coded pricing of pegged collateral is noted as a risk~\cite{zbandut2025curation} and discussed in governance and blog posts~\cite{capo,prevert2024hardcoded,steakhouse2025oracles}, but its losses have not been measured across events. These works give this paper its building blocks.

Works studying vault curation in isolated lending markets treat a depeg as one risk among several~\cite{zbandut2025curation,zbandut2026vault}. Koh and Pukthuanthong model how the recursive pledging of wrapped stablecoins across lending protocols amplifies liquidation cascades~\cite{koh2026wrapped}, the leverage side of the structure whose failure we measure. Wen and Chen reconstruct the November 2025 cascade, scan the Morpho and Euler markets that took the failed coins as collateral, and find the public lending exposure negligible~\cite{wen2026contagion}. In our broader scan the exposure is far larger (\cref{app:stream}). They name the systematic measurement of oracle configurations across lending markets as future work, which this paper supplies.

\section{Model}\label{sec:model}

\subsection{System model}\label{sec:system}

\begin{figure}[t]
 \centering
 \includegraphics[width=\textwidth]{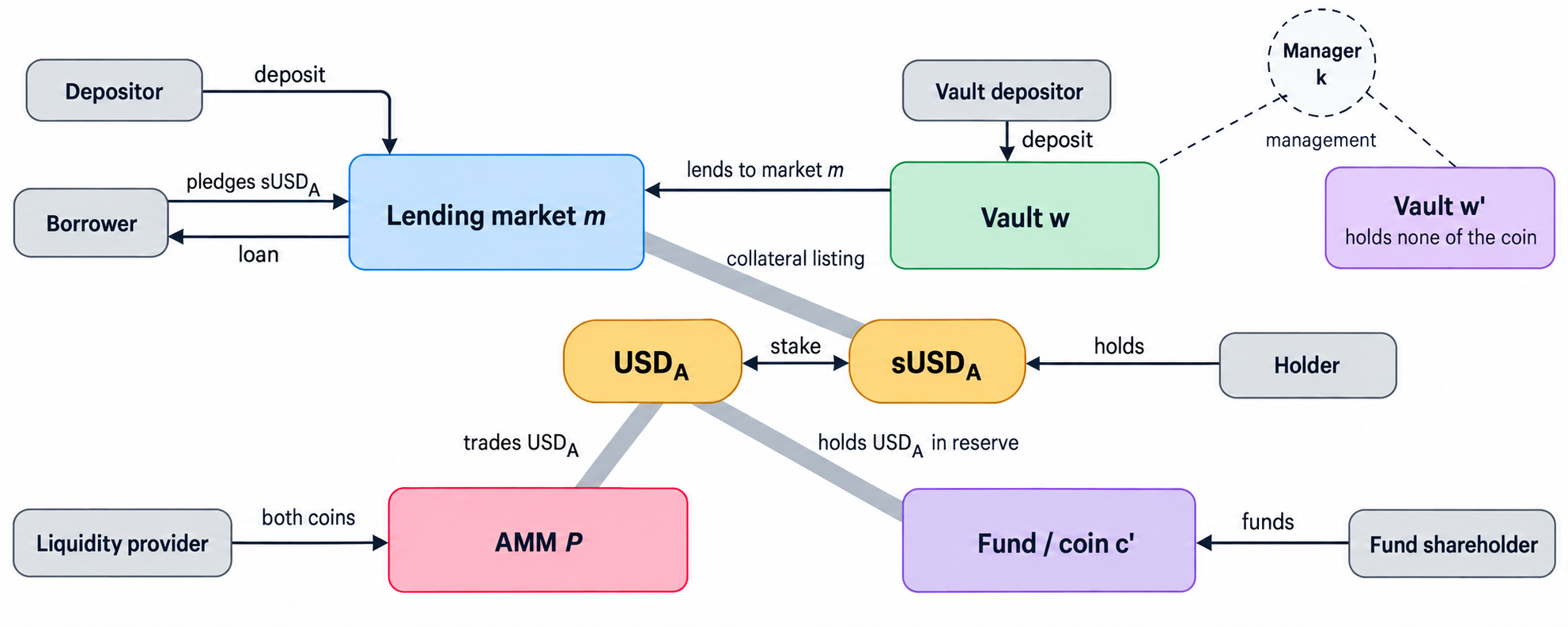}
 \caption{The integration graph of one coin. Solid arrows: money from a party into a place. Grey bands: the edges that carry the coin. Dashed: management, on which no money moves. From any party the arrows lead to the place that holds its money and from there to the coin.}
 \label{fig:system}
\end{figure}

The \emph{integration graph} of a coin $c$ records every place the coin sits and whose money ties to it (\cref{fig:system}). Its nodes are the coin (or staked $c$), the lending markets, vaults, \acp{amm}, funds and other coins that hold it, and the managers of the vaults. Five kinds of edge connect them, and the parties who can lose stand on the edges: holders of $c$ itself, market depositors, vault depositors, borrowers, liquidity providers and fund shareholders.

\vspace{-0.1cm}
\noindent\textit{Listed as collateral.} Lending market $m$ accepts $c$ (or staked $c$) as collateral. The edge carries the oracle price $\booked_m$ and its pricing rule (below), the liquidation threshold $\lltv_m < 1$ and penalty $\beta_m$, the quantity pledged $Q_m$, the supply cap $S_m$, the debt $D_m$ and utilization $u_m$, and the loan liquidity $\liq_m$: the loan asset in the market plus what the allocators of the vaults lending to $m$ can move in. If $m$ is isolated, its bad debt falls on its own depositors; if it is part of a shared pool, on every depositor (\cref{sec:bg-lending}).

\noindent\textit{Allocated by a vault.} Vault $w$ lends $a_{w,m}$ of its deposits to market $m$, within its cap on $m$ and its allocator's limits. The exposure is in every market $w$ lends to that lists $c$.

\noindent\textit{Traded on an \ac{amm}.} \ac{amm} $P$ holds $c$ against a sound (base) coin, with depth $d_P$, the value of the sound side, and its \ac{amm} parameters (e.g., amplification $A$ for Curve).

\noindent\textit{Held by a fund.} A fund or coin $c'$ holds $c$, or positions that hold $c$, as a share $s_{c'}$ of its assets; if $c'$ is itself a pegged coin, the graph continues to expand.

\noindent\textit{Managed together.} Manager $k$ runs vaults $w$ and $w'$; the edge carries no money, and $w'$ may lend to no market that lists $c$. \Cref{sec:vaults} shows why it belongs in the graph.

$H_c$ is the supply valued at par, over every chain the coin is issued on with bridged copies counted once, plus the claims denominated in the coin, such as loans of the coin itself.
Two prices attach to $c$: the market price $p$, what one unit sells for on chain now, and the oracle price $\booked_m$, what market $m$ lends against and liquidates at (\cref{sec:bg-lending}). Par is $\pbar$: $1$ for the coin and the exchange rate for its staked version. A \emph{fixed} rule sets $\booked_m = \pbar$ whatever the market says, and a \emph{market-tracking} rule sets $\booked_m = p$ (\cref{sec:markets}).

\textit{Assumptions}. Every market lends a sound coin unless specified. Every oracle reports according to its rule, so manipulated and stale reports~\cite{eskandari2021sok,mackinga2022twap} are outside the model.

\subsection{Threat model}\label{sec:threat}

The threat is a depeg, followed by whoever can profit from an oracle price that has not moved with it. What and who causes the depeg stays outside our analysis.

A depeg is the market price falling below par; \cref{sec:events} fixes how far and for how long before we count one. Five causes appear in the events we study: a \emph{reserve loss} (Stream xUSD); an \emph{unauthorized mint} of units that nothing backs (Resolv USR, Kelp rsETH); a \emph{panic with reserves intact} (USDC in 2023); a \emph{verification failure}, when the provider attesting to the reserves stops (Main Street msUSD); and a \emph{redemption cut}, when the issuer lowers what a unit redeems for (Usual USD0++). The cause only sets whether the loss is permanent. That turns on the \emph{recovery price} $p_r$: what a unit is worth once the incident is settled, the market price where a pool of size quotes one, otherwise what the issuer pays per unit. While redemption at par stays open, a depositor cannot withdraw but has lost nothing yet. Once it is shut, the loss is real (\cref{sec:total}).

After the depeg, an \emph{arbitrageur} buys $c$ at $p$ on a DEX, pledges them in a lending market whose oracle price is $\booked_m$, borrows up to $\lltv_m\booked_m$ per token, and never repays. We call this trade \emph{oracle arbitrage}, and the loan asset it takes out of a market the \emph{outflow}. Everyone else acts as in \cref{sec:bg-lending}, in its own interest and taking the depeg as given.

For every party $i$ we write $L_i(p)$ for their losses when the coin falls to $p$. \Cref{sec:losses} gives a formula for each $L_i$ and \cref{fig:curves} in \cref{app:proofs} plots them. We show who ends up with the loss, and how much of $L$ can be computed before the event.
\section{Loss Mechanisms}\label{sec:losses}

For each edge of \cref{sec:system} we ask two questions. (i) How much is
the loss? (ii) Who ends up holding the loss? Derivations and proofs are in \cref{app:proofs}.

\subsection{Direct holdings}\label{sec:holders}

Three parties hold the coin itself, so their loss comes directly from the fall in price.

\textit{Holders.} A party holding $H$ units of a token with par value $\pbar$ loses $L_{\mathrm{hold}}(p)$ of \cref{eq:holder_lp} below, the line of \cref{fig:curves}a; for the coin itself $\pbar = 1$. The loss is permanent only if the coin never returns to par or the holder sells before it does.

\textit{Liquidity providers.} Stablecoin \acp{amm} concentrate their liquidity near the peg: Curve's StableSwap through an amplification parameter $A$~\cite{curve}, Uniswap v3 through a price range around par~\cite{uniswapv3}, and Solidly-style pools such as Velodrome through the invariant $x^3y + xy^3 = k$~\cite{velodrome}. Each buys the depegging coin near par until its other side is nearly gone. \Cref{app:pool} solves each \ac{amm}'s composition at the depegging price $p$ from its invariant, and all three end between the same two bounds. Write $V$ for the \ac{amm}'s value at par, twice the depth $d_P$ of \cref{sec:system}. Then \cref{eq:holder_lp} bounds $L_{\mathrm{LP}}(p)$, and the coin side is counted with the holders in \cref{eq:eone}.
\begin{equation}\label{eq:holder_lp}
L_{\mathrm{hold}}(p) = H(\pbar - p), \qquad
1-\sqrt{p} \ \le\ \frac{L_{\mathrm{LP}}(p)}{V} \ \le\ 1-p.
\end{equation}
The lower bound is the \ac{amm} with no concentration; the upper bound is the holder line. The loss rises with the concentration and reaches the upper bound as $A \to \infty$ or the range shrinks to par. \Cref{fig:curves}b plots the StableSwap curve: at $A=100$ it sits within two cents of the holder line for every $p$ below $0.9$. \Cref{sec:empirics} measures the losses per event.

\textit{Fund shareholders.} A fund or coin $c'$ that holds $c$ as a share $s_{c'}$ of its assets loses $L_{c'}(p)/V_{c'} = s_{c'}(1-p)$ per dollar of its value at par $V_{c'}$, shared among its shareholders. If $c'$ is itself a pegged coin it can also depeg, and this section applies again with $c'$.

\subsection{Lending markets}\label{sec:markets}

A depeg also costs the depositors and borrowers of the markets that accept the coin as collateral. Whether borrowers are liquidated in time decides who suffers the losses. A market-tracking oracle lets liquidators act as the price falls, while a fixed oracle does not (\cref{sec:system}). Panels (d) to (g) of \cref{fig:curves} plot the losses. Borrower losses are calculated position by position, with $D$ and $Q$ the debt and collateral of one position; market-level losses sum over positions, and a quantity summed over a market carries the subscript $m$, as in $D_m$ for its total debt. This subsection fixes one market and drops the subscript $m$.

\noindent\textit{Market-tracking oracle: liquidation in time.} A position with debt $D$ against $Q$ units becomes liquidatable once $p < D/(\lltv Q)$ if the oracle tracks the market. A liquidator seizes $(1+\beta)D$ collateral at the oracle price $p_\ell$ of that moment, so the borrower pays \cref{eq:penalty} (\cref{fig:curves}e; a market adds one such step per position). In a panic with intact reserves the coin returns to par, and the borrower also misses the recovery on the collateral taken at the low price: $D((1+\beta)/p_\ell - 1)$ in all (\cref{app:extract}). \Cref{sec:fingerprints} measures these liquidations in the March 2023 USDC event.
\begin{equation}\label{eq:penalty}
L_{\mathrm{bor}} = \beta D \quad \text{once the price touches } \frac{D}{\lltv Q},
\end{equation}
Depositors lose nothing if every position is liquidated before $p$ falls below $(1+\beta)D/Q$, while the collateral still covers the debt and the bonus. A liquidator must sell the seized units to recover what it repaid, so a liquidation pays only while the sale moves the price by less than $\beta/(1+\beta)$ (\cref{app:extract}). When the AMMs are too thin for that (\cref{app:extract}), or the price crosses the line within one block or the chain is congested~\cite{qin2021liquidations}, liquidations stop. An unliquidated position then sits as it would under a fixed rule, and what follows applies to it. Its collateral is sold at $p_r$ in the end, and since every position had $D \le \lltv\pbar Q$ when the depeg began, the depositors of market $m$ lose at most the market-tracking case of \cref{eq:perdollar} below: the same shape as the fixed-rule loss, on the debt already lent.

\noindent\textit{Fixed rule: nothing is liquidated.} The oracle reports par whatever the market price, $\booked = \pbar$, so no position becomes liquidatable as the price falls. Above $\lltv\booked$ nothing happens inside the market: every position's collateral still covers its debt, and buying the coin to pledge it does not pay. Borrowers lose only as holders, $Q(\pbar - p)$ on the units they pledged (\cref{eq:holder_lp}), and depositors lose nothing.

Below $\lltv\booked$ the oracle arbitrage of \cref{sec:threat} will happen: buy at $p$, pledge, borrow $\lltv\booked$ per unit, and never repay, for a gain of $\lltv\booked - p - \kappa$ per unit, where $\kappa$ is the gas cost of such an arbitrage. New loans against the coin should therefore appear once $p$ falls below $\lltv\booked - \kappa$, which \cref{sec:empirics} tests. Units created without backing (e.g., an unauthorized mint) cost nothing, so the loans can come before the price falls (\cref{sec:total}, case E2).

Two caps bound the outflow: the market's loan asset $\liq$ and the collateral's supply cap $S$. The pledged units stay unliquidated, since a liquidator would receive units valued at $\booked$ that sell for $p$. They are sold at $p_r$ after the depositors' liquidation below, and the depositors lose the outflow less the sale's return (\cref{eq:extract}). Positions open before the fall with debt above $pQ$ are abandoned too and add at most $D_m\pos{1 - p_r/(\lltv\booked)}$ (\cref{app:extract}). The market is now at full utilization: no one can borrow or withdraw.
\begin{equation}\label{eq:extract}
X = \min(\liq,\ \lltv\booked S), \qquad L_{\mathrm{dep}} = X\pos{1 - \frac{p_r}{\lltv\booked}}.
\end{equation}
Write $\delta_m(p)$ for the most a depeg to price $p$ can cost the depositors of market $m$ per dollar of its deposits $T_m$, so that $L_{\mathrm{dep}} \le T_m\,\delta_m(p)$:
\begin{equation}\label{eq:perdollar}
\delta_m(p) =
\begin{cases}
\pos{1 - p/(\lltv\booked)} & \text{fixed oracle,}\\[2pt]
u_m\,\pos{1 - p/((1+\beta)\lltv\pbar)} & \text{market-tracking oracle, liquidations fail}
\end{cases}
\end{equation}
with $p_r = p$ and $u_m = D_m/T_m$ the market's utilization. Under a fixed oracle every dollar, lent or idle, can end behind an abandoned loan; under a market-tracking one only the lent share is at risk, and only if liquidation fails. \Cref{fig:curves}d draws both curves.

What a depositor can still do is trade its claim for collateral. In one transaction it repays an abandoned position's debt, receives the position's collateral at the oracle price, $(1+\beta)/\booked$ units per dollar repaid, and withdraws the same amount from its own deposit. The collateral received is worth $(1+\beta)p_r/\booked$ per dollar of claim.

\noindent\textit{The interest rate.} The analysis so far held every debt fixed. Once the market sits at full utilization the borrow rate surges: under Morpho's adaptive rate model~\cite{morphoirm} it settles at 800\% per year, where debt doubles about monthly (\cref{fig:curves}f). Other protocols' fixed top rates are in \cref{app:rate}. Three things follow.

(A) The \emph{borrowers} still in the market are trapped. A borrower who borrowed $D_0$ against collateral now worth $pQ$ owes $D(t)$ and repays only while that is worth it, so its loss beyond the principal is:
\begin{equation}\label{eq:trap}
L_{\mathrm{bor}}(t) = \min\big(pQ,\ D(t)\big) - D_0, \qquad D(t) = D_0\exp\!\Big(\int_0^t r(s)\,\dd s\Big).
\end{equation}
It grows with the debt until $D(t)$ reaches $pQ$, from which the borrower will not repay (repaying is net negative). For example, a borrower whose collateral was worth twice the principal crosses it in two months (\cref{fig:curves}g). This is where the loss lands on the loopers of \cref{sec:bg-lending}. The paid interest is a transfer to the depositors of the same market.

(B) Meanwhile the \emph{depositors} who stay earn the same rate on their deposit shares. A depositor can use its shares at any time to liquidate an abandoned position, as above, and take collateral at the oracle price, or it can wait for the shares to grow. But the debts grow at the same speed. After a position's debt exceeds what its collateral covers at the oracle price, $\booked Q/(1+\beta)$, whoever liquidates first takes the collateral, and once it is gone the market holds nothing but bad debt. \Cref{app:queue} solves this \emph{timing game}: every depositor liquidates at the moment the debt reaches that value, when the collateral is exactly exhausted, and receives $p_r/(\lltv\booked)$ per dollar lent, its pro-rata share. Depositors who act late leave their share to those who act on time. Each abandoned position has its own such moment; the arbitrage positions, all at the threshold $\lltv$, share one.

(C) The protocol's literal debt, meanwhile, overstates the loss. After a position is abandoned no further fund leaves the protocol, and the literal loss splits into two parts:
\begin{equation}\label{eq:phantom}
D(t) - p_r Q \;=\; \underbrace{(D_0 - p_r Q)}_{\text{real loss}} \;+\; \underbrace{(D(t) - D_0)}_{\text{phantom interest}}.
\end{equation}
The real loss is fixed at the loan, and the \emph{phantom interest} was never paid and grows without bound. Reading the loss off the ledger overstates it by that amount, so \cref{sec:empirics} values debt at the oracle price of the loan and collateral at the recovery price, and reports the phantom interest beside the real loss.

The market stays pinned until borrowers repay or the bad debt is written off, and until then the depositors who did not convert hold a claim they cannot use. No balance sheet records this as a loss. We measure it in dollar-days, the withheld balance integrated over the time it is withheld, and keep it apart from the other losses (\cref{sec:total}).

\subsection{Loss contagion across markets}\label{sec:vaults}

\Cref{sec:markets} gives one market's loss. The loss then travels: to the depositors who stay, to the vaults lending to the market, to the markets sharing a vault with it, and to other vaults under the same manager. Who holds it is decided by the order in which depositors leave, since the remaining liquidity pays those who leave first in full, so a \emph{run} follows. Let a market or vault hold deposits $T$, an unwritten loss $U$ and idle liquidity $I$, what the outflow left in the damaged market or the unlent money in a vault's other markets. The first $I$ dollars leave whole and those who stay hold the whole loss:
\begin{equation}\label{eq:exit}
\text{loss per dollar} =
\begin{cases}
\min\big(1,\ U/(T-I)\big) & \text{for a depositor who stays,}\\
0 & \text{for one who left.}
\end{cases}
\end{equation}
The pro-rata figure $U/T$ is the place's average and no one's loss; everything that lowers $I$ moves loss onto those who stay. Below we detail each contagion path.

\noindent\textit{The depositors who stay.} An isolated market keeps the loss with its own depositors. A shared pool spreads one collateral's loss over every depositor of the pool, including those who never touched the coin, and the outflow pins the pool's loan asset, so every borrower of that asset pays the rate of \cref{sec:markets} whatever its collateral. A coin listed on several chains carries the loss to each of them, as rsETH did on two (\cref{sec:empirics}).

\noindent\textit{The vaults that lend to the market.} A vault $w$ stands in its allocated markets as a depositor, so a depeg to price $p$ costs it at most
\begin{equation}\label{eq:vault}
L_w \le T_w \sum_{m} a_{w,m}\,\delta_m(p),
\end{equation}
with $a_{w,m}$ the allocation to $m$ and $\delta_m$ of \cref{eq:perdollar}. \Cref{eq:exit} says who holds it.

Two things lower the vault's idle liquidity $I$. (i) The public allocator (\cref{sec:bg-lending}), which anyone can call, moves the vault's idle balance, within the caps the manager set, into a market that asks for more liquidity. An arbitrageur can pull it in and borrow it in one transaction, raising the outflow there (\cref{eq:extract}). (ii) Withdrawals are paid from the vault's sound markets, so a large enough wave pins a sound market $m'$ the vault also lends to that holds none of the coin (\cref{app:pinned}).

\noindent\textit{The sound markets that share a vault with it.} A sound market pinned by those withdrawals is at full utilization, so \cref{eq:trap} applies to its borrowers: their collateral has nothing to do with the coin, yet they pay the surging interest, and its depositors who stayed are frozen. The freeze ends by itself while the collateral stays healthy. The rate pushes the borrowers to repay or into liquidation, and each repayment releases deposits. The cost is the rate they pay and the time depositors wait, not bad debt. If that collateral is itself a pegged token, the forced selling may push it off its peg: borrowers sell it to repay, liquidators sell what they seize, and holders rush to redeem. \Cref{sec:markets} then applies to it too. In the msUSD event the AVLT market that shared a vault with the damaged one was run in turn, and AVLT then depegged (\cref{app:msusd}).

The freeze can \emph{cascade}: a pinned market freezes the money other vaults allocated to it, their withdrawals pin their other markets: damaged market $\to$ vault $\to$ shared market $\to$ vault. Each hop moves a share of the previous one and stops where idle liquidity absorbs the wave. The cascade spreads the freeze, not the bad debt (\cref{app:pinned}), though the frozen money can exceed the bad debt that started it (\cref{sec:empirics}).

\noindent\textit{Other vaults with the same manager.} Depositors also flee vaults that merely look related. Let manager $k$ run vault $w$, which holds the damaged market, and vault $w'$, which lends to no market listing the coin. The depositors of $w'$ cannot confirm the scope of the loss but know their manager is stuck in $w$. \Cref{eq:exit} applies to $w'$ with $U$ the loss they expect, and their withdrawals pin the sound markets $w'$ lends to. In the msUSD event every vault under the same manager lost deposits, \nCuratorRunFamilyDrawdownPct{} of the family's TVL within 10 days, while only one of them was exposed~\cite{todayindefi2026alphaping} (\cref{sec:fingerprints}).

\subsection{The worst-case loss bound}\label{sec:total}

Summing the losses of \cref{sec:holders,sec:markets,sec:vaults} over the coin's reach, defined below, gives the total loss. A claim denominated in the coin, such as a loan of the coin itself, counts with the holders; a vault's depositors count with the lenders.
\begin{equation}\label{eq:ledger}
\text{total loss} = \text{holders} + \text{\acp{amm}} + \text{lenders} + \text{borrowers}.
\end{equation}

Each loss carries one of three properties. A loss is \emph{destroyed} when the value is gone for everyone: the loss of backing reserve, or an unauthorized mint not backed by reserves. It is \emph{transferred} when it is in someone else's pocket: the arbitrageur's gain, a liquidator's bonus, a panic seller's loss to a buyer who holds to re-peg. It is \emph{withheld} when the money exists but cannot be reached, such as a frozen deposit. Two events with the same total loss can differ completely: in a panic with intact reserves every dollar one party lost is in another party's pocket, while a reserve loss destroys at least what the reserves lost.
Which terms a depeg opens, and with which property, follows from its type through the recovery price $p_r$ of \cref{sec:threat}; \cref{tab:activation} in \cref{app:proofs} lists the cases.

The exposure behind every term of \cref{eq:ledger} is on-chain state, so its worst value can be computed before the depeg, over the coin's \emph{reach}: markets that list $c$ or its staked version, vaults lending to such a market or managed by the same manager, \acp{amm} holding $c$, and coins with $c$ in reserve. The reach tells where a fund loss can travel. How far the freeze and the rate surge travel along the chain of vaults and markets depends on their depositors. These are outside the bounds below. Two cases bound the worst: (A) the price falls to zero with the supply fixed, a reserve loss or panic that never re-pegs; (B) units can be created without limit, an unauthorized mint.

In (A), the total loss in \cref{eq:ledger} is at most
\begin{equation}\label{eq:eone}
E_1(c,t) = H_c + E^{\mathrm{cred}}_1(c,t) + \sum_{m} \beta_m D_m + \sum_{P} d_P + \sum_{c'} E_1(c',t),
\end{equation}
with $m$ over the markets in the reach, $c'$ over the coins holding $c$ in reserve, which we assume form no cycle, and the credit term $E^{\mathrm{cred}}_1(c,t) = \sum_{m} \min(\liq_m,\ \lltv_m\booked_m Q_m)$, which also bounds the loan assets that leave those markets. In (B) the loan assets that leave the markets are at most
\begin{equation}\label{eq:etwo}
E_2(c,t) = \sum_{m} \min\!\big(\liq_m,\ \lltv_m\booked_m S_m\big),
\end{equation}
with $S_m = \infty$ when market has no cap; every listing market counts, whatever its oracle, because the minted tokens are pledged while the price is still at par. The total loss in (B) is at most \cref{eq:eone} with $E_2$ in place of the credit term: the holders of the genuine units lose at most $H_c$ however many units are minted. \Cref{app:bound} derives both, and the price form $E_1(c,t;p)$ used in \cref{sec:backtest}.

\section{Empirical Analysis}\label{sec:empirics}

This section answers three research questions with on-chain data. \emph{RQ1.} Which coins depegged, and who lost how much? (\cref{sec:events}) \emph{RQ2.} Do the losses arise as \cref{sec:losses} derives? (\cref{sec:fingerprints}) \emph{RQ3.} Does the pre-event bound of \cref{sec:total} hold on the events, and how much is exposed today, according to our estimation? (\cref{sec:backtest})

\noindent\textit{Setup.}\label{sec:data} We take every stablecoin or pegged token listed by DefiLlama \cite{defillama} or CoinGecko \cite{coingecko}, their DEX pools, and \nCensusProtocols{} lending protocols on \nCensusUniverseChains{} chains. Pools and markets are scanned from \nCensusWindowStart{} to \nCensusWindowEnd{}, with prices read from \acp{amm} at \nCensusPriceGridHours{}-hour steps in the lending census and \nPcPriceGridHours{}-hour steps around each fall in the price census. An event's onset is the start of the first six hours in which the coin's price stays below $0.97$ of par; the recovery prices behind the loss figures are read on \nValuationDate{} (Detailed in \Cref{app:census}).

\subsection{Depeg events and losses}\label{sec:events}

Across \nPcInScope{} stablecoins and pegged tokens over \nPcThreshold{} market cap, \nPcUnionSizedFamilies{} coin families depegged in the window, \nPcUnionShareOfSizePct{} (\nPcUnionPeakSizeUsd{}) of the peak market cap scanned. \cref{tab:pricecensus} lists them, and the coin holders alone lost at least \nPcHolderPeakUsd{} at the lowest prices, \nPcHolderRealizedUsd{} of it still standing at the valuation date.
In \nCensusEvents{} events, lending markets kept pricing the coin above the market as it fell, found across \nCensusProtocols{} protocols. These cost \nDedupTotal{} at the valuation date after deduplication: \nCreditChannel{} (\nDedupCreditPct{}) fell on depositors and coin lenders, \nDedupBorrower{} (\nDedupBorrowerPct{}) on borrowers, \nDedupPools{} (\nDedupPoolsPct{}) on \acp{amm}, and \nDedupFreeFloat{} (\nDedupFreeFloatPct{}) on holders of the free float. Of the total, \nNovelOtherDedupUsd{} comes from the events outside the five reconstructed below, estimated from market state without position replay, an upper bound (\cref{tab:ledger}).

The share inside lending markets and \acp{amm} runs from \nPartXusdEmbPct{} for xUSD and \nPartUsrEmbPct{} for USR to \nPartRsethEmbPct{} for rsETH (\cref{tab:partition}). Five events carry \nDedupTotalFive{} of the total, \nDepositorSideFive{} of it on depositors. \cref{tab:ledger} gives their losses. Each is summarized below (\cref{app:events}).

\noindent\textit{1. rsETH, \nRsethOnset{} (unauthorized mint).} A forged bridge message released \nRsethForgedUnits{} rsETH, pledged in lending markets, and \nRsethAttacker{} was borrowed against them. Governance lowered the oracle and liquidated the positions. The pools' deficit has not been written down. Holders lost nothing, as the backing is restored (\cref{app:rseth}).

\noindent\textit{2. USR, \nUsrOnset{} (unauthorized mint).} Tokens minted with a compromised key were sold on and pledged in markets that priced USR at par, and \nUsrExtraction{} was borrowed against them. Most of the loss fell on holders and on those who had pledged USR (\cref{tab:ledger}). The issuer later settled USR at par for pre-incident holders and at \nUsrRecoveryPrice{} per unit after, and RLP at \nRlpRecoveryPrice{}~\cite{resolv2026postmortem,resolv2026recovery} (\cref{app:usr}). We value the loss at the compensation price.

\noindent\textit{3. Stream xUSD, \nStreamOnset{} (reserve loss).} The reserve manager reported a loss, starting the depeg. Depositors lost \nStreamDepositor{}, most of it on Plume and Plasma. On Plume one address borrowed heavily against minted xUSD (\cref{sec:fingerprints}; reports relate this to the issuer~\cite{blockeden2025stream}). Holders lost \nStreamFreeFloatLoss{}, and lenders withdrew \nStreamFlight{} (\cref{app:stream}).

\noindent\textit{4. msUSD, \nMsusdOnset{} (verification failure).} The provider attesting to the reserves ended its coverage~\cite{protos2026msusd} and the price fell to \nMsusdMinPrice{}. Depositors lost \nMsusdDepositor{} and holders \nMsusdFreeFloatLoss{}; lenders withdrew \nMsusdFlight{}, a run that spread to the manager's other vaults and to other tokens, AVLT and AZND (\cref{sec:fingerprints}, \cref{app:msusd}).

\noindent\textit{5. USD0++, \nUsdZeroPPOnset{} (redemption cut).} The issuer set the early-redemption value below par~\cite{usual2025dualexit} and the price fell to \nUsdZeroPPMinPrice{}. Oracle arbitrage is barely enabled at this price (\cref{app:usd0pp}), and holders, borrowers and \acp{amm} carry the loss (\cref{tab:ledger}). This panicked the lenders, who withdrew \nUsdZeroPPFlight{}, the largest flight of the five.

\begin{table}[tbp]
\caption{The loss by place for the major depeg events, \nCensusWindowStart{} to \nCensusWindowEnd{}, in USD on \nValuationDate{}. Every unit counted once by where it sat at the onset (\cref{tab:partition}). \emph{Depositors}: principal borrowed against the coin, not returned. \emph{Coin lenders}: loss of those who lent the coin itself. \emph{Borrowers}: net value drop of positions using the coin as collateral. \emph{Holders}: drop on units outside lending markets and \acsp{amm}. \emph{\acsp{amm}}: liquidity providers' loss. \emph{Frozen}: balance withheld $\times$ days withheld, not a loss, outside every total. \emph{--}: not measured.}
\label{tab:ledger}
\centering
\scriptsize
\setlength{\tabcolsep}{2pt}
\begin{tabular}{@{}l r r r r r r r@{}}
\toprule
& \multicolumn{3}{c}{\textbf{Lending}} & \multicolumn{2}{c}{\textbf{Direct holding}} & & \\
\cmidrule(lr){2-4}\cmidrule(lr){5-6}
\textbf{Event (type)} & \textbf{Depositors} & \textbf{Coin lenders} & \textbf{Borrowers} & \textbf{Holders} & \textbf{\acsp{amm}} & \textbf{Frozen (M\$$\cdot$d)} & \textbf{Total} \\
\midrule
rsETH (unauthorized mint) & \nRsethDepositor{} & \nRsethLoanSide{} & \nRsethBorrowerLoss{} & \nRsethFreeFloatLoss{} & \nRsethDedupLp{} & -- & \nRsethDedup{} \\
USR (unauthorized mint) & \nUsrDepositor{} & \nUsrLoanSide{} & \nUsrBorrowerLoss{} & \nUsrFreeFloatLoss{} & \nUsrDedupLp{} & \nUsrFrozenMDays{} & \nUsrDedup{} \\
xUSD (reserve loss) & \nStreamDepositor{} & \nStreamLoanSide{} & \nStreamBorrowerLoss{} & \nStreamFreeFloatLoss{} & \nStreamDedupLp{} & \nStreamFrozenMDays{} & \nStreamDedup{} \\
msUSD (verification failure) & \nMsusdDepositor{} & \nMsusdLoanSide{} & \nMsusdBorrowerLoss{} & \nMsusdFreeFloatLoss{} & \nMsusdDedupLp{} & \nMsusdFrozenMDays{} & \nMsusdDedup{} \\
USD0++ (redemption cut) & \nUsdZeroPPDepositor{} & \nUsdZeroPPLoanSide{} & \nUsdZeroPPBorrowerLoss{} & \nUsdZeroPPFreeFloatLoss{} & \nUsdZeroPPDedupLp{} & \nUsdZeroPPFrozenMDays{} & \nUsdZeroPPDedup{} \\
Other events & \nNovelOtherDepositorUsd{} & \nNovelOtherLoss{} & -- & \nNovelOtherHolderUsd{} & \nNovelOtherAmmUsd{} & -- & \nNovelOtherDedupUsd{} \\
\midrule
\textbf{All} & \textbf{\nDepositorSide{}} & \textbf{\nLoanSideAll{}} & \textbf{\nDedupBorrower{}} & \textbf{\nDedupFreeFloat{}} & \textbf{\nDedupPools{}} & \nFrozenTotalMDays{} & \textbf{\nDedupTotal{}} \\
\bottomrule
\end{tabular}
\end{table}

No single split holds across events (\cref{tab:ledger}): USR put most of its loss on holders, xUSD most of it on depositors, rsETH all of it on depositors who never held the coin. \Cref{sec:markets} places each event: the pricing rule opened the depositor channel in four of the five (besides USD0++). Outside the losses ledger, lenders withdrew \nFlightTotal{} during the five windows (\cref{sec:fingerprints}), and the markets book \nPhantomInterestThree{} of phantom interest at pinned rates that no one will pay (\cref{eq:phantom}).

\subsection{Mechanism signatures}\label{sec:fingerprints}

This section answers RQ2: we test each loss mechanism of \cref{sec:losses}, except the holders' and \ac{amm} losses (\cref{sec:holders}), which are the price gap times the amount held.

\noindent\textit{Which party the oracle rule exposes.} In March 2023 USDC fell to about \nUsdcGauntletLow{} and Aave's oracle followed it~\cite{gauntlet2023usdc}, so no one borrowed against a fixed price. Instead borrowers were liquidated at the depressed price, and the liquidations left about \nUsdcGauntletBadDebt{} of bad debt~\cite{gauntlet2023usdc}. In the USR event the oracles held par, no position could be liquidated, and depositors lost \nUsrDepositor{}. Each oracle rule protects one party and exposes the other.

\begin{figure}[t]
 \centering
 \includegraphics[width=\textwidth]{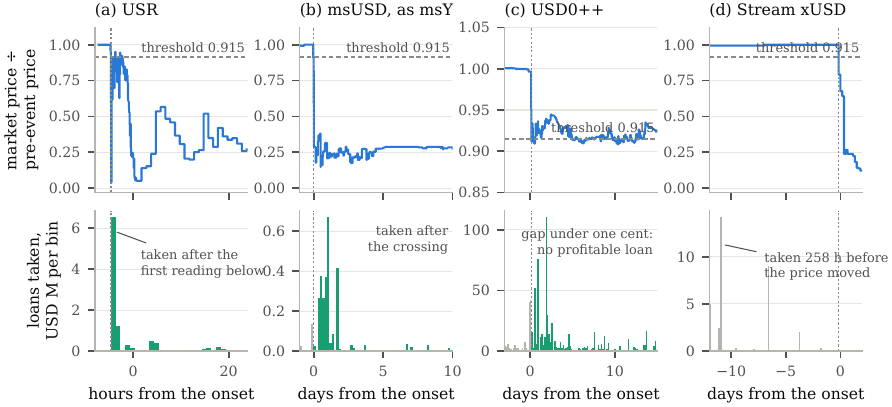}
 \caption{Top: market price relative to par. Dashed: the highest funded liquidation threshold. Dotted: the first reading below it. Bottom: loans taken against the collateral, dark after that reading.}
\label{fig:timing}
\end{figure}

\noindent\textit{When the loans were taken.} \Cref{sec:markets} predicts new loans against the coin only once its price is below $\lltv\booked$ by more than the slippage of buying it. \Cref{fig:timing} checks this on four events. \emph{USR}: the price first fell below the line at \nUsrCrossingUtc{} UTC and loans came after it. \emph{msUSD}: \nMsusdBorrowAfterCrossingPct{} of the loans came after the crossing, though little was left to borrow in a market already at full utilization (\cref{app:msusd}). \emph{USD0++}: the price crossed the line but stayed within \nUsdZeroPPCrossingDepthPct{} of it, less than the slippage on trades above \nKappaBreakevenSize{}, and all but \nUsdZeroPPExtraction{} of the loans were repaid or remain collateralized (\cref{app:usd0pp}). \emph{Stream}: the loans came before the crossing, because of the suspicious borrowing (\cref{app:stream}).

\noindent\textit{Which caps bound.} Of the \nCapMarketsWithOutflow{} markets that carried the five events' outflow, a cap filled during the event on \nCapMarketsBound{}. Nearly all of it in one market: Aave's rsETH supply cap of 530K had \nCapRsethMainnetRoomPct{} of room at the snapshot, the mint filled it at \nCapRsethMainnetFilledUtc{}. Elsewhere no cap bound: Morpho markets have none, and the vaults' caps stayed above the liquidity available, so liquidity and new deposits set the outflow.

\begin{figure}[t]
 \centering
 \begin{minipage}[t]{0.44\textwidth}
  \centering
  \includegraphics[width=\textwidth]{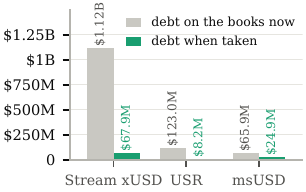}
  \caption{Debt of the abandoned positions when taken and on the books now, for the three events with any. The gap is the phantom interest of \cref{eq:phantom}, \nPhantomInterestThree{} in all.}
  \label{fig:phantom}
 \end{minipage}\hfill
 \begin{minipage}[t]{0.54\textwidth}
  \centering
  \includegraphics[width=\textwidth]{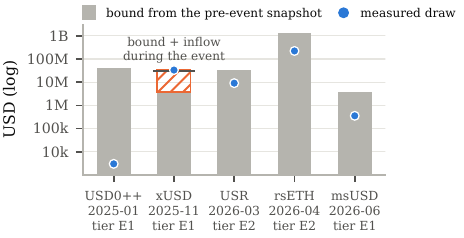}
  \caption{Pre-event bound (bar, $E^{\mathrm{cred}}_1$ or $E_2$ of \cref{sec:total} by depeg type) against the measured outflow (point). Hatched: the excess. Tick: bound plus liquidity deposited during the event.}
  \label{fig:bound}
 \end{minipage}
\end{figure}

\noindent\textit{Phantom interest.} By \cref{eq:phantom}, once a market is pinned, the debt on its books keeps growing by interest no one will pay, while the loss stops growing. The abandoned positions on the replayed markets owed \nStreamAbandonedDebtOnset{} when taken in Stream, \nUsrAbandonedDebtOnset{} in USR and \nMsusdAbandonedDebtOnset{} in msUSD. At the valuation date the same positions carry \nStreamAbandonedDebtNow{}, \nUsrAbandonedDebtNow{} and \nMsusdAbandonedDebtNow{} (\cref{fig:phantom}). rsETH and USD0++ had no abandoned position: rsETH's were liquidated within days of the mint (\cref{app:rseth}), and USD0++ never fell far enough for the arbitrage to pay. Across the three events this phantom interest totals \nPhantomInterestThree{}.

\noindent\textit{Frozen deposits and the pinned rate.} In \nFixedPointMarkets{} markets on \nFixedPointEvents{} events the borrow rate reached the \nPeakAprPct{} cap, the fixed point of \cref{sec:markets}. Markets on a fixed rate curve stop at the top of their curve instead (xUSD of \cref{tab:frozen}). Across the \nFrozenEvents{} census events measured, \nPinnedEvents{} had a pinned market, and \nFrozenTotalMDays{} million dollar-days of deposits were frozen.

\begin{table}[tbp]
\caption{Frozen deposits and the pinned rate. Debt pinned: borrow balance of the largest pinned market at the window's end. Hours: longest run at or above 99\% utilization while the token was below $0.97$. At cap: markets whose rate hit the 800\% cap. Frozen: balance withheld $\times$ days withheld. AVLT: the market of \cref{fig:run}, which held no msUSD.}
\label{tab:frozen}
\centering
\scriptsize
\setlength{\tabcolsep}{4pt}
\begin{tabular}{@{}l l r r r r r r@{}}
\toprule
\textbf{Event} & \textbf{Token} & \textbf{Markets} & \textbf{Debt pinned} & \textbf{Hours} & \textbf{Peak rate} & \textbf{At cap} & \textbf{Frozen (M\$$\cdot$d)} \\
\midrule
USR & USR & \nTfourUsrMarkets{} & \nTfourUsrStock{} & \nTfourUsrHours{} & \nTfourUsrApr{} & \nTfourUsrCap{} & \nTfourUsrMDays{} \\
 & wstUSR & \nTfourWstUsrMarkets{} & \nTfourWstUsrStock{} & \nTfourWstUsrHours{} & \nTfourWstUsrApr{} & \nTfourWstUsrCap{} & \nTfourWstUsrMDays{} \\
 & RLP & \nTfourRlpMarkets{} & \nTfourRlpStock{} & \nTfourRlpHours{} & \nTfourRlpApr{} & \nTfourRlpCap{} & \nTfourRlpMDays{} \\
msUSD & msUSD & \nTfourMsusdMarkets{} & \nTfourMsusdStock{} & \nTfourMsusdHours{} & \nTfourMsusdApr{} & \nTfourMsusdCap{} & \nTfourMsusdMDays{} \\
 & AVLT & \nTfourAvltMarkets{} & \nTfourAvltStock{} & \nTfourAvltHours{} & \nTfourAvltApr{} & \nTfourAvltCap{} & \nTfourAvltMDays{} \\
Stream & xUSD & \nTfourXusdMarkets{} & \nTfourXusdStock{} & \nTfourXusdHours{} & \nTfourXusdApr{} & \nTfourXusdCap{} & \nTfourXusdMDays{} \\
USD0++ & USD0++ & \nTfourUsdZeroPPMarkets{} & \nTfourUsdZeroPPStock{} & \nTfourUsdZeroPPHours{} & \nTfourUsdZeroPPApr{} & \nTfourUsdZeroPPCap{} & \nTfourUsdZeroPPMDays{} \\
rsETH & rsETH & \nTfourRsethMarkets{} & \nTfourRsethStock{} & \nTfourRsethHours{} & \nTfourRsethApr{} & \nTfourRsethCap{} & \nTfourRsethMDays{} \\
\midrule
\multicolumn{5}{@{}l}{All \nFrozenEvents{} census events measured} & \nPeakAprPct{} & \nFixedPointMarkets{} & \nFrozenTotalMDays{} \\
\bottomrule
\end{tabular}
\end{table}

\begin{figure}[t]
 \centering
 \includegraphics[width=0.98\textwidth]{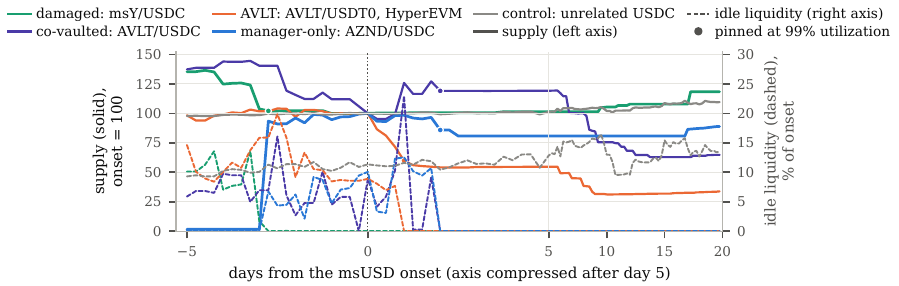}
 \caption{The msUSD run reaches markets that never held the coin. Solid: supply, onset = 100. Dashed: idle liquidity, share of supply. Dots: Full utilization. Time compressed after day 5.}
 \label{fig:run}
\end{figure}

\noindent\textit{Deposit runs.} \Cref{fig:run} follows the msUSD event along the cascade path of \cref{sec:vaults}. At the onset the co-vaulted market and the control held \nFigFiveCoVaultedIdleOnsetPct{} and \nFigFiveCtrlIdleOnsetPct{} of their supply idle. Once pinned, every repayment is withdrawn at once and supply falls, while idle liquidity stays zero. The damaged market pinned \nMsusdLiquidityOnsetLeadH{} hours before the depeg, as lenders withdrew \nMsusdPreOnsetWithdrawn{}. Its supply then grew by \nFigFiveCoAfterPinGrowthPct{} with idle liquidity at zero because of phantom interest. \emph{Shared vault}: the AVLT market funded by the same vault pinned at \nFigFiveCoVaultedPinnedHours{} hours and deposits fell to \nFigFiveCoVaultedTroughPct{}. \emph{Same manager}: the AZND market, which never held the coin, pinned at \nFigFiveManagerCleanPinnedHours{} hours, followed by idle liquidity withdrawn. \emph{Second order}: the AVLT market on \nFigFiveCuratorChain{}, its collateral under a run of its own, pinned at \nFigFiveAvltPinnedHours{} hours and fell to \nFigFiveCuratorTroughPct{}. The control market's idle liquidity never fell below \nFigFiveCtrlIdleMinPct{}.

\subsection{The loss bound}\label{sec:backtest}

We apply the bounds of \cref{sec:total} to the five events and the stablecoins today. For the two mint events the bound is $E_2$; for the other three it is $E^{\mathrm{cred}}_1$.

The total bound of \cref{eq:eone} holds in all five events (\cref{app:totalbound}): \nUsrTotalBound{} against a loss of \nUsrTotalBoundRealized{} for USR, \nStreamTotalBound{} against \nStreamTotalBoundRealized{} for Stream, \nMsusdTotalBound{} against \nMsusdTotalBoundRealized{} for msUSD, and more than tenfold margins for rsETH and USD0++. Most of each bound is the holders' term. The credit term, the part that market configuration sets, holds in four events (\cref{fig:bound}): USR's outflow of \nUsrExtraction{} sits under \nBoundSelUsr{}, nearly all of it routable from vaults, and rsETH used \nBoundSelUseRsethPct{} of the \nBoundSelRseth{} the pools could lend against it. The breach is Stream, \nStreamExtraction{} against \nBoundSelStream{}, because \nInflowStream{} of new deposits reached the pinned markets during the event. Adding them leaves \nUncoveredAfterInflowStream{} uncovered. The bound is a snapshot, and new inflow can raise the loss (\cref{sec:discussion}).

The per-vault sum works the same way. The Gauntlet USDC Frontier vault held \nGauntletTotalAssets{} at the USR snapshot with \nGauntletCommitted{} of caps on markets pricing USR at par, so a depositor could have read that at most \nGauntletMaxLossPct{} of the vault was exposed (\nGauntletRealizedLossPct{} realized).

Run today, \cref{tab:live} gives the exposure of the \nLiveBoundCoins{} pegged coins that \nLiveBoundProtocols{} lending protocols list on \nLiveBoundChains{} chains as of \nLiveBoundAsOfDate{}, each coin its own scenario. Among the \nLiveBoundRowsListed{} largest, a fall to half of par would cost from \nLiveBoundRankListedLastAtFifty{} for \nLiveBoundRankListedLastCoin{} to \nLiveBoundTopOneAtFifty{} for \nLiveBoundTopOneCoin{}, most of it on holders. The lenders' term counts only fixed-oracle markets at half of par and is largest for \nLiveBoundLenderTopOneCoin{} (\nLiveBoundLenderTopOneValue{}), \nLiveBoundLenderTopTwoCoin{} (\nLiveBoundLenderTopTwoValue{}) and \nLiveBoundLenderTopThreeCoin{} (\nLiveBoundLenderTopThreeValue{}). At zero every market counts and it reaches \nLiveBoundLenderZeroTopOneValue{} for \nLiveBoundLenderZeroTopOneCoin{} (\cref{app:fulltable}).

\section{Discussion}\label{sec:discussion}

\noindent\textit{Defenses.} Each term of the bounds in \cref{sec:total} is a parameter someone controls, so a defense is a cap on one term. \textit{The supply term}: a cap or a timelock on minting. \textit{The pricing rule}: neither rule avoids the loss, it chooses who bears it. A market-tracking oracle liquidates in time but, in a panic, liquidates positions that were never at risk (USDC in 2023). A fixed oracle spares borrowers but opens the oracle arbitrage (\cref{sec:markets}), whose steps are ordinary actions, so only an immediate pause stops it. \textit{The lending term}: the market's supply cap and the allocator's flow caps set the outflow. The supply cap on rsETH stopped the mint outflow within minutes of filling (\cref{sec:fingerprints,sec:backtest}).

\noindent\textit{After the depeg.} The model prices each loss at the recovery price of \cref{sec:threat}, the market price where liquidity exists or what the issuer pays per unit. A payment beyond that moves the loss between parties and does not remove it. Resolv settled USR and RLP claims at stated rates, which the ledger takes as their recovery prices (\cref{app:usr})~\cite{resolv2026recovery}. In the rsETH event the seized tokens restored the issuer's backing while the pools' deficit stands (\cref{app:rseth}). No compensation has followed for Stream, and msUSD's issuer bought back part of the supply from its treasury.

\noindent\textit{Why it recurs.} After these losses, \nLiveBoundProtocols{} lending protocols still price pegged collateral by a fixed rule, with up to \nLiveBoundLenderTopOneValue{} of loans on one coin open to a fall to half of par (\cref{app:fulltable}). The rule persists because in normal times it costs nothing and spares borrowers a liquidation on every dip, while the exposure it creates falls on depositors who cannot see it. The bound of \cref{sec:total} computes it in advance from public state.

\noindent\textit{Limitations.} (1) Losses that never touch a chain are invisible to us. Stream's off-chain manager loss enters only through its on-chain consequences. (2) The bound is a statement about snapshot state. It does not predict the timing or size of the runs of \cref{sec:vaults}. (3) The exposure that stands today (\cref{tab:live}) is each coin's own worst case, not a joint scenario. (4) The census and the bound need a market price, so a coin with little \ac{amm} depth is read poorly and is visible only through its lending markets (\cref{app:pricecensus}).

\section{Conclusion}\label{sec:conclusion}

When a pegged coin breaks its peg, the loss lands where the coin was integrated, not only where it was held. We modeled the places a coin reaches as a graph, derived each party's loss as a function of the post-depeg price, and bounded the worst case from public chain state. Over three years, \nCensusEvents{} depegs in which lending markets kept pricing the coin above the market cost \nDedupTotal{}, \nDedupCreditPct{} of it borne by lenders who never held the coin, and the total bound held on every event we reconstructed. Since every input is public, the same computation can be run for any listed coin before its peg breaks.

\bibliographystyle{splncs04}
\bibliography{refs}

\clearpage
\appendix
\crefalias{section}{appendix}
\crefalias{subsection}{appendix}

\section{Derivations and Proofs}\label{app:proofs}

\begin{table}[H]
\caption{Which loss channels each depeg type opens, and with which property. Lenders lose under a fixed oracle once $p < \lltv\booked$ (\cref{eq:extract}) and under a market-tracking one only if liquidations fail (\cref{eq:perdollar}). Borrowers also lose as holders on the pledged units they paid for (\cref{sec:markets}).}
\label{tab:activation}
\footnotesize
\setlength{\tabcolsep}{4pt}
\begin{tabularx}{\textwidth}{@{}>{\raggedright\arraybackslash}p{2.1cm} >{\raggedright\arraybackslash\hsize=0.95\hsize}X >{\raggedright\arraybackslash\hsize=1.15\hsize}X >{\raggedright\arraybackslash\hsize=0.95\hsize}X >{\raggedright\arraybackslash\hsize=0.95\hsize}X@{}}
\toprule
\textbf{Depeg type} & \textbf{Holders, \acsp{amm}} & \textbf{Lenders} & \textbf{Borrowers} & \textbf{Runs} \\
\midrule
Reserve loss & destroyed & real, $p_r$ below par & rate trap & all paths \\
\addlinespace
Unauthorized mint & diluted until burned & real if unbacked, else frozen & rate trap & every listing chain \\
\addlinespace
Panic, reserves intact & transferred to buyers & frozen, $p_r$ back to par & liquidated if tracked & ends at re-peg \\
\addlinespace
Verification failure & marked down meanwhile & real or frozen, open & rate trap & open \\
\addlinespace
Redemption cut & marked down to the issuer's new price & real only below $\lltv\booked$; none in USD0++ & rate trap & lender flight \\
\bottomrule
\end{tabularx}
\end{table}

\begin{figure}[htbp]
 \centering
 \includegraphics[width=\textwidth]{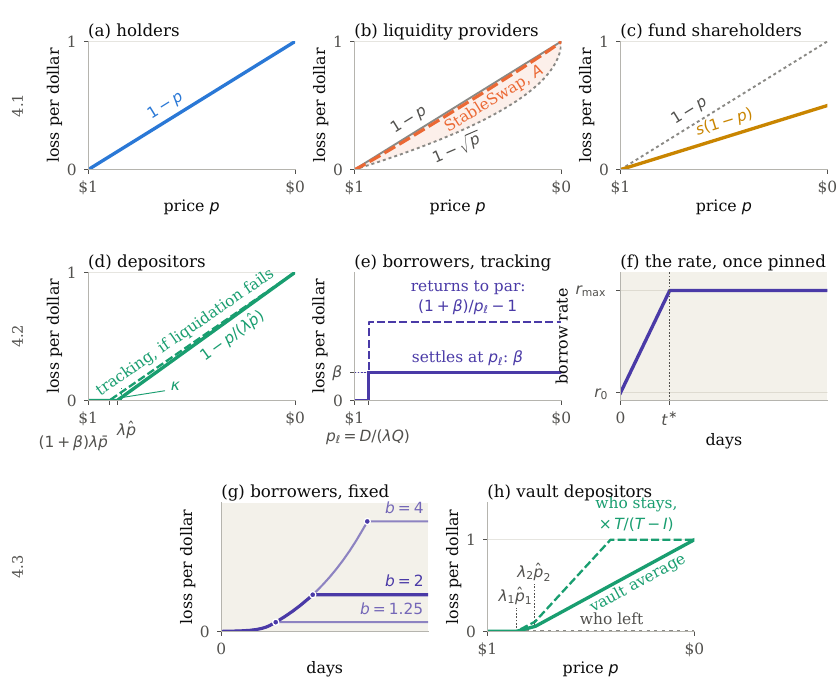}
 \caption{The loss curves from the model alone, as loss per dollar put in. Row 1, holders: (a) a holder, (b) a StableSwap \acs{amm} between its two bounds, (c) a fund holding the coin as a share $s$ of its assets. Row 2, one lending position: (d) a depositor under a fixed price (solid) and under a market-tracking price when liquidations fail (dashed, \cref{eq:perdollar}); (e) a borrower under a market-tracking price, the penalty at the lowest price (solid) or the loss if the coin returns to par (dashed); (f) the borrow rate once utilization is pinned; (g) a borrower paying that rate under a fixed price, collateral worth $b$ times the principal. Row 3: (h) a vault depositor across two markets, before and after the exit order of \cref{eq:exit}. (f) and (g) run over days at full utilization. Constants: $\lltv\booked = 0.86$, $\kappa = 0.02$, $\beta = 4.4\%$, $A = 100$, $s = 0.5$, rate model of \cref{app:rate}.}
 \label{fig:curves}
\end{figure}

\subsection{The \acs{amm} loss curves}\label{app:pool}

\begin{figure}[t]
 \centering
 \includegraphics[width=0.72\textwidth]{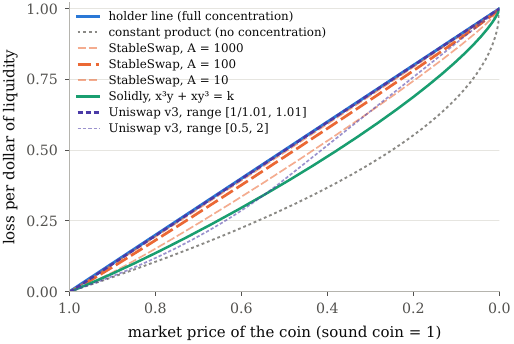}
 \caption{Loss per dollar of liquidity for the three \acs{amm} families of
 \cref{sec:holders}, against the two bounds: the holder line (full
 concentration) and the constant-product line (none). StableSwap at
 $A = 10$, $100$ and $1000$; the Solidly invariant; Uniswap v3 positions
 over $[1/1.01, 1.01]$ and $[0.5, 2]$. Model only; no on-chain data
 enters.}
 \label{fig:amm}
\end{figure}

An \ac{amm} holds a balance $x$ of the coin and $y$ of the sound coin and
quotes the marginal price $-\dd y/\dd x$ along its invariant. We open
each position at par with one dollar of value, push the coin in until the
quoted price equals the target $p < 1$, and value the position at
$px + y$; the loss per dollar is $1 - (px + y)$. Fees are ignored. Two
cases bracket the rest. Under the constant-product invariant $xy = 1/4$
the price is $y/x$, so $x = 1/(2\sqrt{p})$, $y = \sqrt{p}/2$, and the
loss is $1 - \sqrt{p}$. An \ac{amm} that quotes par until its sound side is
gone sells all of $y$ at par for any $p < 1$ and is then worth $p$, a
loss of $1 - p$. \Cref{fig:amm} plots the curves below against these two
bounds.

\noindent\textit{StableSwap.}
A two-asset StableSwap \ac{amm} with balances $x$ and $y$ and amplification
$A$ satisfies, in the whitepaper's convention~\cite{curve},
\[
4A(x+y) + D = 4AD + \frac{D^3}{4xy},
\]
where $D$ is the invariant and equals $x+y$ when the balances are equal.
We normalize $D = 1$, so a balanced \ac{amm} holds half a unit of each coin
and is worth one dollar at par. Multiplying by $y$ gives a quadratic in
$y$, $4A y^2 + (4Ax + D - 4AD)\,y - D^3/(4x) = 0$, whose positive root
gives $y(x)$. The marginal price of the coin, by implicit
differentiation, is
\[
p(x) = \frac{4A + D^3/(4x^2y)}{4A + D^3/(4xy^2)},
\]
equal to $1$ at $x = y$. As sellers push the coin into the \ac{amm}, $x$ rises
and $p(x)$ falls. For a target price $p < 1$ we solve $p(x) = p$ for $x$
by bracketing (the function is monotone), read $y(x)$, and value the \ac{amm}
at $px + y$. The two limits are immediate. At $A = 0$ the invariant
reduces to $xy = D^2/4$, the constant-product rule. As $A \to \infty$ it
reduces to $x + y = D$, so the \ac{amm} quotes par regardless of its
composition and sellers take all of $y$ at par. For finite $A$ the loss lies between the two and rises with $A$ (\cref{fig:amm}).

\noindent\textit{Uniswap v3.}
A position with liquidity $L$ over the range $[p_a, p_b]$, $p_a < 1 <
p_b$, holds $x = L(1/\sqrt{P} - 1/\sqrt{p_b})$ of the coin and $y =
L(\sqrt{P} - \sqrt{p_a})$ of the sound coin at a price $P$ inside the
range~\cite{uniswapv3}; below $p_a$ it holds $x = L(1/\sqrt{p_a} -
1/\sqrt{p_b})$ and no sound coin. Opened at par it is worth $L(2 -
1/\sqrt{p_b} - \sqrt{p_a})$, so the loss per dollar at $p$ is
\[
1 - \frac{2\sqrt{p} - p/\sqrt{p_b} - \sqrt{p_a}}{2 - 1/\sqrt{p_b} - \sqrt{p_a}}
\ \ (p_a \le p < 1), \qquad
1 - p\,\frac{1/\sqrt{p_a} - 1/\sqrt{p_b}}{2 - 1/\sqrt{p_b} - \sqrt{p_a}}
\ \ (p < p_a).
\]
For a range symmetric in $\sqrt{P}$, $p_a = 1/p_b = 1/r$, the position
holds equal dollar amounts at par and the value per dollar is $(2\sqrt{p}
- (p+1)/\sqrt{r})/(2 - 2/\sqrt{r})$ inside the range and $p(\sqrt{r} +
1)/2$ below it. Inside the range the value is at most $\sqrt{p}$ because
$(1 - \sqrt{p})^2 \ge 0$, and at least $p$ because $p \ge 1/r$ there;
below the range $p(\sqrt{r}+1)/2$ lies between $p$ and $\sqrt{p}$ because
$\sqrt{p} \le 1/\sqrt{r}$. The curve therefore sits between the two
bounds, and below the range its value rises with $r$, so the loss rises as the range narrows, and as $r \to 1$ it tends to $1 - p$.

\noindent\textit{Solidly.}
The invariant $x^3y + xy^3 = k$ of Velodrome and Aerodrome stable
pools~\cite{velodrome}, opened balanced with $x = y = 1/2$ so that $k =
1/8$, has marginal price $(3x^2y + y^3)/(x^3 + 3xy^2)$, equal to $1$ at
$x = y$. It has no parameter: its concentration is fixed. We solve $p(x)
= p$ by bracketing as for StableSwap, with $y(x)$ the positive root of
the cubic $xy^3 + x^3y = k$. The curve lies between the two bounds, near par about as concentrated as StableSwap at $A = 10$ (\cref{fig:amm}).

\subsection{Derivation of \cref{eq:extract}}\label{app:extract}

\emph{Threshold.} Repaying debt $D_0$ frees collateral worth $p Q_0$, a
gain of $p Q_0 - D_0$, positive exactly when $p Q_0 \ge D_0$. A position
the market admits has $D_0 \le \lltv\booked Q_0$, so $p \ge \lltv\booked$
implies $p Q_0 \ge D_0$ for every position and every loan is repaid. A
position opened at $p_b$ and abandoned has inflow $\lltv\booked$ per unit
and outflow $p_b + \kappa$ per unit, so its gain is positive exactly when
$p_b < \lltv\booked - \kappa$. On $\lltv\booked - \kappa \le p_b <
\lltv\booked$ the entry cost blocks new positions but does not reach the
stock: a position with $p Q_0 < D_0$ gains by abandoning there too.
Such a position leaves depositors $D_0 - p_r Q_0 \le D_0\big(1 - p_r/(\lltv\booked)\big)$,
since $D_0 \le \lltv\booked Q_0$; summed over the positions open before the fall
this is at most $D_m\pos{1 - p_r/(\lltv\booked)}$.
Under the market-tracking rule the oracle price equals $p$ at every
block, which turns the entry condition into $p \ge \lltv p - \kappa$, which holds for every $p \ge 0$ since
$\lltv < 1$, so no position is opened to be abandoned. For the stock,
take a position with debt $D$ against $Q$ units, liquidated in full at
price $p_\ell$: the liquidator repays $D$ and takes $(1+\beta)D/p_\ell$
units, which requires $p_\ell \ge (1+\beta)D/Q$. The range of prices at
which this is possible, $(1+\beta)D/Q \le p_\ell < D/(\lltv Q)$, is
non-empty because $\lltv(1+\beta) < 1$, which Morpho's bonus formula
guarantees. The borrower started with $Q$ units at par, $D$ in cash and
debt $D$, and ends with $Q - (1+\beta)D/p_\ell$ units, the cash and no
debt; at a settling price $p$ its loss is
$Q(1-p) + D\big((1+\beta)p/p_\ell - 1\big)$, the holder line plus
\cref{eq:penalty} when $p = p_\ell$, and $D((1+\beta)/p_\ell - 1)$ when
the coin returns to par. With a partial liquidation, $D$ is the amount
repaid. A liquidator that repays $r$ receives $(1+\beta)r/p_\ell$ units
and gains if their average sale price is at least $p_\ell/(1+\beta)$,
that is, if the sale moves the price by at most $\beta/(1+\beta)$. If
$p_\ell < (1+\beta)D/Q$, the liquidator takes all $Q$ units for
$p_\ell Q/(1+\beta)$ of debt, and depositors keep a shortfall
$D - p_\ell Q/(1+\beta)$; a position never liquidated and sold at $p_r$
leaves $D - p_r Q$, which is smaller. Summing the positive parts over
positions, and using $D \le \lltv\pbar Q$, which holds for every position
because each was healthy when the depeg began, at an oracle price no
higher than par, gives the market-tracking case of \cref{eq:perdollar}. The market totals
$D_m - p_r Q_m/(1+\beta)$ do not give the bound: the surplus of one
position returns to its own borrower and covers no other shortfall.

\emph{Amount.} With $p_b < \lltv\booked - \kappa$ the abandoned position
has a strictly positive gain per unit, so positions are opened until a
constraint binds. The loan asset that can leave is at most the liquidity
$\liq$, and the credit the market extends is at most $\lltv\booked$ per
unit on at most $S$ units, so $X = \min(\liq, \lltv\booked S)$. The
departed loans are backed by $Q' = X/(\lltv\booked)$ units, which the
market sells at $p_r$, recovering $p_r Q'$. Depositors lose
$X - p_r Q' = Q'(\lltv\booked - p_r) = X(1 - p_r/(\lltv\booked))$, which
is \cref{eq:extract}.

\emph{Split.} The openers gain $(\lltv\booked - p_b - \kappa) Q'$.
Subtracting from $L_{\mathrm{dep}}$ gives $Q'(p_b - p_r) + \kappa Q' \ge 0$ since
$p_r \le p_b$. This residual is paid by depositors and received by no
one.

\subsection{The adaptive rate model}\label{app:rate}

Morpho markets price borrowing with an adaptive
curve (\texttt{AdaptiveCurveIRM}) \cite{morphoirm}. Write $u$ for
utilization, $u^\ast = 0.9$ for
the target, and $r^\ast(t)$ for the rate-at-target. The instantaneous borrow
rate is $r = \gamma(u)\,r^\ast(t)$, with $\gamma(u) = 1 + 3\,(u - u^\ast)/(1 -
u^\ast)$ above the target, so $\gamma(1) = 4$. The rate-at-target adapts
with the utilization error $e(u) = (u - u^\ast)/(1 - u^\ast)$ at speed
$50$ per year: over an interval of length $\Delta t$ with constant $u$,
$r^\ast$ is multiplied by $\exp(50\, e(u)\, \Delta t)$, and is clamped to
$[0.1\%, 200\%]$ per year. At $u = 1$ the error is $1$, so
$r^\ast(t) = \min(r^\ast_0 e^{50t}, 2)$ with $t$ in years and the borrow
rate is $4 r^\ast(t)$. The doubling time of the rate is $\ln 2/50$ years,
$5.06$ days. From the model's initial $r^\ast_0 = 4\%$ the cap is reached
at $t^\ast = \ln 50/50$ years, $28.6$ days, after which the borrow rate is
$8$ per year. Interest accrues by compounding, so a debt held a full year
at the cap grows by the factor $e^{8} = 2{,}980.96$, an annual yield of
$297{,}996\%$. The cumulative rate is
\[
\int_0^t r(s)\,\dd s =
\begin{cases}
\dfrac{4 r^\ast_0}{50}\big(e^{50t} - 1\big) & t \le t^\ast,\\[6pt]
\dfrac{4 r^\ast_0}{50}\big(e^{50t^\ast} - 1\big) + 8\,(t - t^\ast) & t > t^\ast,
\end{cases}
\]
which equals $0.157$ at $t^\ast$: the debt grows by $17\%$ during the ramp,
then doubles every $\ln 2/8$ years, $31.6$ days. A borrower whose
collateral is worth $b$ times the principal at the market price abandons on
the first $t$ with $\int_0^t r \ge \ln b$; for $b = 1.25, 2, 4$ this is day
$32$, $53$ and $85$, the points marked in \cref{fig:curves}g.

Aave v3~\cite{aavev3}, Compound v3~\cite{compoundv3} and Euler v2~\cite{eulerv2} use linear-kink curves instead: the borrow rate rises with utilization at one slope up to a target utilization and at a steeper slope above it, and at 100\% utilization it stands at the top of the curve, the base rate plus the two slopes, a per-market parameter read from chain state. That rate is reached as soon as utilization is pinned and does not grow with time, so debt in such a market compounds at a fixed rate below the adaptive fixed point. \Cref{sec:empirics} reports the peaks observed in each event.

\subsection{The pinned sound market}\label{app:pinned}

Let a vault $w$ hold a share $a_{w,m}$ of its assets in the damaged
market $m$ and a share $a_{w,m'}$ in a sound market $m'$ with utilization
$u_{m'}$ and debt $D_{m'}$. Withdrawals cannot be paid from $m$, so each
withdrawn dollar takes $a_{w,m'}/(1-a_{w,m})$ from $m'$. A withdrawal wave
of size $W$ raises the utilization of $m'$ to
$D_{m'}/(D_{m'}/u_{m'} - W a_{w,m'}/(1-a_{w,m}))$, and $m'$ is pinned at
full utilization once $W$ exceeds its idle liquidity divided by
$a_{w,m'}/(1-a_{w,m})$. The cascade iterates on the integration graph. Start with the damaged market pinned. At each step, every vault with a positive allocation in a pinned market is run: its depositors withdraw up to its idle liquidity, and each withdrawn dollar is drawn from its unpinned markets in proportion to their allocations, $a_{w,m''}/(1 - a_{w,P})$ for market $m''$, where $a_{w,P}$ is the vault's total allocation to pinned markets. A market joins the pinned set when the withdrawals it receives, summed over vaults, exceed its idle liquidity. The set stops growing when no unpinned market receives more than it can pay, which happens because each hop moves at most the deposits the running vaults hold in the next market, a share of a share, and every market absorbs the first $I$ dollars of any wave.

\subsection{Who gets the collateral}\label{app:queue}

Consider a market at full utilization whose borrowers have all abandoned their positions, with $Q$ units of collateral in total, debt $D(t) = D_0\,g(t)$, and depositors' claims that grow by the same factor $g(t)$, since at full utilization the supply rate equals the borrow rate less a protocol fee we ignore. The market price satisfies $p < \booked/(1+\beta)$, so no outside liquidator acts. A depositor that repays $d$ receives $(1+\beta)d/\booked$ units, so the collateral absorbs at most $K = \booked Q/(1+\beta)$ of repayment, and a converted dollar is worth $v = (1+\beta)p_r/\booked$. The collateral seizable against the debt is $\min(Q,\ (1+\beta)D(t)/\booked)$ units. Before the time $t^\ast$ at which $D(t^\ast) = K$, a full repayment leaves units with the borrower; after it, $K < D(t)$ and repayments are served first come, first served until the collateral is gone. Each depositor $i$ chooses when to convert its claim $c_i(t) = c_i(0)\,g(t)$ and receives $v\,c_i(\tau_i)$ on the part that is served. Converting before $t^\ast$ forfeits growth and secures nothing that is not also available at $t^\ast$: the claims at $t^\ast$ total $D(t^\ast) = K$, so every depositor is served then, whether or not others converted earlier. A depositor who plans to convert at some $T > t^\ast$ is served only if capacity remains, and gains by moving to just before $T$; this unravels every such plan back to $t^\ast$. So all depositors convert at $t^\ast$, the collateral is exactly exhausted, and each receives $v\,c_i(0)\,g(t^\ast) = c_i(0)\,p_r/(\lltv\booked)$, its pro-rata share of $Q$ per dollar first lent; the phantom interest of \cref{eq:phantom} on both sides of the book cancels. Depositors who act late, out of inattention or because their claims are small, leave their share to those who act at $t^\ast$. With $g$ from \cref{app:rate}, $t^\ast$ is the day $D(t)/D_0$ reaches $1/((1+\beta)\lltv)$, a few weeks at the pinned rate.

With many abandoned positions the same holds position by position. Position $j$ has collateral $Q_j$, initial debt $D_j(0)$ and its own $t^\ast_j$, at which its collateral is exhausted by claims that were worth $D_j(0)$ at the start. A converted dollar buys the same $(1+\beta)/\booked$ units in any position, so the collateral of position $j$ ends up with claims equal to its initial debt, $Q_j/D_j(0)$ units per dollar. Positions with more collateral per dollar of debt have later $t^\ast_j$ and are worth more to whoever converts into them; which depositors get them is not pinned down, since any assignment of claims to positions is an equilibrium and a depositor who deviates towards a better position risks being served nowhere. The arbitrage positions, all at the threshold $\lltv$, share one $t^\ast$ and carry most of the abandoned debt in the events of \cref{sec:empirics}. Borrowers who still service their positions repay in cash, which depositors withdraw at par in the order of \cref{eq:exit}.

\subsection{Derivation of \cref{eq:phantom}}\label{app:phantom}

The principal $D_0$ left the protocol when the loan was taken; the collateral $Q$ stayed and is worth $p_r Q$ when sold. No later event
moves money: the borrower is gone, and a liquidation exchanges the debt
claim for the same $Q$ units. So the real loss is $D_0 - p_r Q$ at every
$t$. At full utilization no repayment and no withdrawal alters the
balance, so the debt compounds at the instantaneous rate and
$D(t) = D_0\exp(\int_0^t r)$. With $r(s) \ge r_{\min} > 0$ the integral
diverges and so does $D(t)$. Adding and subtracting $D_0$ in the booked
shortfall $D(t) - p_r Q$ gives \cref{eq:phantom}. Any figure computed
from $D(t)$, including a protocol's or an indexer's bad-debt field,
carries the term $D(t) - D_0$; the real loss $Q(\lltv\booked - p_r)$
contains no rate and no time.

\subsection{Derivation of the bounds}\label{app:bound}

\emph{The bound \cref{eq:eone}.} Bound each term of \cref{eq:ledger} by the exposure standing at $t$. Holders, including borrowers' pledged units, the units supplied to markets that lend $c$, and the coin side of every \ac{amm}, lose at most $H_c$, the par value of the supply over every chain the coin is issued on, bridged copies counted once, plus the claims denominated in the coin; this counts pledged and \ac{amm} units once each, and counting them again in the market and \ac{amm} terms only loosens the bound. Holder losses are counted per unit and charged to the unit's holder at the onset, as in \cref{tab:partition}, so trades among holders net out. A loan of the coin itself is a claim denominated in the coin and is counted in $H_c$, which therefore bounds the lender's loss as well. Lenders of market $m$ lose at most what can be borrowed against the pledged collateral, $\lltv_m\booked_m Q_m$, and at most the loan asset available, $\liq_m$ (\cref{eq:extract} with $p_r = 0$); the minimum holds for every pricing rule since $\booked_m$ is read at $t$, and $E^{\mathrm{cred}}_1$ sums it over the markets. The same minimum bounds the loan assets that leave $m$, which is what is borrowed against that collateral. Liquidity providers of \ac{amm} $P$ lose at most the sound side $d_P$ that can be drained against the coin (\cref{app:pool}, $p = 0$). A coin $c'$ holding $c$ in reserve is marked down and re-enters as the shocked coin; the bound $E_1(c',t)$ drops the reserve share, which is at most one, and acyclicity makes the recursion finite. Borrowers lose at most the liquidation penalty on their debt, $\beta_m D_m$ per market, whether the liquidation is triggered by the price or by the rate of \cref{sec:vaults}; interest they pay at the pinned rate is a transfer to the depositors of the same market. Positions open at the onset add no term: a position abandoned under a fixed rule, or liquidated late under a tracking rule, costs its borrower and the depositors together the holder loss on its units plus at most the liquidator's bonus, which is below $\beta D$ since $pQ/(1+\beta) < D$ in that case, and both are already counted. Compensation paid afterwards takes over a loss already counted in one term, so it moves the loss between parties and leaves the total where it was. For the price form, replace $H_c$ by $H_c(1-p)$, each lender term by $\min(\liq_m, \lltv_m\booked_m Q_m)\pos{1 - p/(\lltv_m\booked_m)}$ (\cref{eq:extract} with $p_r = p$), and each \ac{amm} term by $d_P$ times the \ac{amm}'s loss per dollar at $p$ from \cref{app:pool}, which bounds the loss on its sound side, the providers' loss beyond the drop on the coin side already in $H_c(1-p)$; each replacement bounds the edge's loss at price $p$, each is non-increasing in $p$, the lender terms vanish for $p \ge \lltv_m\booked_m$, and at $p = 0$ the sum is $E_1(c,t)$. Write $E_1(c,t;p)$ for the sum so replaced, so that $E_1(c,t;0) = E_1(c,t)$. For a coin that returns above the price at which positions were liquidated, $p$ is the lowest price reached: the borrower's loss beyond $\beta D$, $(1+\beta)D(1/p_\ell - 1)$, is at most the holder term on the units sold at $p_\ell \ge p$. \emph{The bound \cref{eq:etwo}.} Fix a market $m$ that lists $c$. The outflow cannot exceed the liquidity $\liq_m$, and cannot exceed the credit on the most collateral the market accepts, $\lltv_m\booked_m S_m$, hence the minimum. Every pricing rule contributes. The created units are pledged while the price is still at par, so $\booked_m$ is par under a market-tracking rule as well as under a fixed one, and the market lends $\lltv_m\booked_m$ per unit either way; a fall in the price afterwards only lowers what a later pledge would raise. The rule decides what happens next, whether liquidators can act as the price falls (\cref{sec:markets}), not how much leaves. Under unlimited creation the collateral costs nothing at the margin, so no term involves the creator's endowment, and outflows across markets add because each market's constraints bind separately. The bound is attained in the frictionless case: create and pledge $\min(S_m, \liq_m/(\lltv_m\booked_m))$ units in each qualifying market and borrow to the limit. \emph{The vault bound \cref{eq:vault}.} A vault's loss per dollar of its deposits is the allocation-weighted sum of the per-market losses per dollar lent; how that loss is shared among its depositors is set by the order in which they leave (\cref{sec:vaults}), and the sum is the vault's average. A market with a fixed rule loses at most $1 - p/(\lltv_m\booked_m)$ per dollar lent once $p < \lltv_m\booked_m$ and nothing above it, if every dollar the vault placed there is borrowed out; a market with a market-tracking rule never crosses the threshold of \cref{sec:markets}, and loses at most \cref{eq:perdollar} on the share $u_m$ of its deposits already lent, with $p_r = p$. Summing the two terms with the allocation weights gives \cref{eq:vault}, which is zero when every market the vault lends to lists none of $c$.

\section{The Events in Detail}\label{app:events}

\subsection{Where the supply sat}\label{app:partition}

One number frames the events: how much of each coin's supply sat inside the places of \cref{sec:system} when its depeg began. For each family token we rebuild the full holder map at the onset block from its transfer events, check it against the token's total supply, and classify every holder by on-chain probes: the lending protocols of the census, the \acp{amm} of the price layer, bridge lockboxes, wrappers and vaults, and plain wallets, with units inside a wrapper attributed through it. \Cref{tab:partition} gives the result; in the de-duplicated column of \cref{tab:ledger} \acp{amm} are charged their measured loss where one exists and units times drop where none does, and borrowers are charged the fall on pledged units whose positions were kept plus the over-collateralization forfeited on abandoned positions, \nDedupBorrowerEquity{} in all; the depositor column takes the rest of the fall on abandoned collateral. Units created after the snapshot, or issued by the issuer to itself, that nobody ever paid for are charged to no party while they stay with their creator, in a wallet or as its own deposit in a lending market, or with whoever received them for free; \nDedupSelfCreatedUsd{} is left out on this rule. The share of the supply that sat inside lending markets and \acp{amm} runs from \nPartXusdEmbPct{} for xUSD and \nPartUsrEmbPct{} for USR, most of whose supply was in a bridge lockbox, to \nPartRsethEmbPct{} for rsETH, and these embedded units are where the losses of this section come from.

\begin{table}[tbp]
\caption{Where each coin's supply sat at its onset block. Lending: pledged
or lent inside the covered protocols. \acsp{amm}: on automated market makers. Both:
their sum. The window's price drop is charged to whoever held each unit
at the onset (\cref{sec:data}).}
\label{tab:partition}
\centering
\footnotesize
\setlength{\tabcolsep}{5pt}
\begin{tabular}{@{}l r r r l@{}}
\toprule
\textbf{Token} & \textbf{Lending} & \textbf{\acsp{amm}} & \textbf{Both} & \\
\midrule
rsETH & \nPartRsethLendPct{} & \nPartRsethDexPct{} & \nPartRsethEmbPct{} & Aave's deposit token alone: \nPartRsethAavePct{} \\
RLP & \nPartRlpLendPct{} & \nPartRlpDexPct{} & \nPartRlpEmbPct{} & \\
msUSD & \nPartMsusdLendPct{} & \nPartMsusdDexPct{} & \nPartMsusdEmbPct{} & \\
USD0++ & \nPartUsdZeroPPLendPct{} & \nPartUsdZeroPPDexPct{} & \nPartUsdZeroPPEmbPct{} & \\
xUSD & \nPartXusdLendPct{} & \nPartXusdDexPct{} & \nPartXusdEmbPct{} & \\
USR & \nPartUsrLendPct{} & \nPartUsrDexPct{} & \nPartUsrEmbPct{} & \nPartUsrBridgePct{} in a bridge lockbox (other chains) \\
\bottomrule
\end{tabular}
\end{table}

\subsection{rsETH}\label{app:rseth}

Units minted through a forged cross-chain message were pledged on Aave on mainnet and Arbitrum and on Spark. The seven addresses named in the pools' incident report~\cite{aave2026rseth} borrowed \nRsethAttacker{} against them, \nRsethAttackerWeth{} WETH and \nRsethAttackerWsteth{} wstETH at the onset price of \$\nRsethEthOnsetPrice{} per ETH; the report's own count is \nRsethOfficialWeth{} WETH and \nRsethOfficialWsteth{} wstETH. On \nRsethAipDate{} the pools' governance lowered the rsETH oracle and liquidated the minter's positions (proposal AIP 478), seizing \nRsethSeized{} rsETH and repaying \nRsethLiqRepaid{} of debt. The pools booked a deficit of \nRsethDeficitWeth{} WETH and \nRsethDeficitWsteth{} wstETH, \nRsethNetLoss{} at the onset price, which stands in full at our valuation block. The seized rsETH went into the issuer's bridge lockbox or was burned, which restored the backing of rsETH held on other chains and returned no WETH to the pools' depositors. Holders lost nothing lasting: the exchange rate recovered once the minted units were burned.

\subsection{USR}\label{app:usr}

Units minted with a compromised issuer key were sold on and pledged by their buyers in markets that priced USR at par and its staked version by the exchange rate. \nUsrExtraction{} of stablecoins were borrowed against them, \nUsrExtractionMainnetMorpho{} of it on mainnet Morpho; the surface is \nUsrSurfaceMarkets{} markets on \nUsrSurfaceChains{} chains, \nUsrSurfaceWrappedPct{} of which list not USR but wstUSR, the issuer's second token RLP, or a principal token on one of these. Depositors lost \nUsrDepositor{} of principal, \nUsrSummaryDepositorUsd{} of it on replayed Morpho positions, \nUsrPtDepositorUsd{} on a market taking the family's principal token and \nUsrStateDepositorUsd{} on Euler and Silo; lenders of USR itself lost \nUsrLoanSide{} more; the drop on every unit of USR and of RLP, a second token of the same issuer, was \nUsrHolder{} at the recovery prices, \nUsrFreeFloatLoss{} of it on the free float that \cref{tab:ledger} charges to holders and \nUsrSelfCreatedHolderUsd{} of it on the attacker's own units, which is not charged; \acp{amm} lost \nUsrLp{}. The issuer later burned the minted units it recovered and settled holders' claims at the rates given below~\cite{resolv2026postmortem,resolv2026recovery}; \cref{sec:discussion} traces what the settlement moved and what it left in place.

\noindent\textit{Timeline.}
The minted units were pledged and \nUsrExtraction{} borrowed on \nUsrOnset{}, the largest hour at \nUsrPeakBorrowLeadHours{} hours before
the mainnet market price crossed the highest funded threshold at
\nUsrCrossingUtc{} UTC (\cref{fig:timing}a); \nUsrBorrowAfterCrossingPct{}
of the window's loan volume came after the crossing. Of the outflow, \nUsrExtractionMainnetMorpho{} was on mainnet Morpho and \nUsrExtractionBase{}
on Base. Lenders withdrew \nUsrFlight{} in the window; the markets then
pinned, and \cref{tab:frozen} gives the freeze on the USR, wstUSR and RLP
families.

\noindent\textit{Recovery price.}
The recovery price for \cref{tab:ledger} is read at \nValuationDate{} under the validity rule of the price census (\cref{app:pricecensus}). No pool trading USR, wstUSR or RLP passes it at the valuation block, so the issuer's settlement, paid through a claim window still open at that date, is what a unit is worth: pre-incident USR and wstUSR at par, units acquired after the incident at \nUsrRecoveryPrice{} per USR, and RLP at \nRlpRecoveryPrice{} in USDC plus RESOLV tokens that we do not value~\cite{resolv2026recovery}. The ledger values every USR unit at the lower rate, since units held before and after the incident are not told apart at the onset, and wstUSR at its conversion rate times that.

\noindent\textit{Caps, liquidity and the allocator.}
Two mainnet markets priced USR at par: the RLP market, with \nUsrCappedHeadroom{} of headroom and \nUsrCappedExtraction{} borrowed, and the USR market, with minted collateral and \nUsrUncappedExtraction{} borrowed; the vault caps of neither bound. At the snapshot the USR-family markets held \nUsrSnapLiquidity{} of loan liquidity against \nUsrSnapLtvHeadroom{} of headroom, and the public allocator could route \nRoutableUsr{} more from vaults that had the markets enabled. During the event \nInflowUsr{} of loan liquidity arrived, so \nUsrInflowSharePct{} of what was borrowed came from liquidity that was not in the markets at the snapshot; the routable term is what lets the pre-event bound of \nBoundSelUsr{} cover the outflow (\cref{fig:bound}).

\subsection{Stream xUSD}\label{app:stream}

The outside manager of the reserves reported a loss and the price went to \nStreamMinPrice{} from a booked value of \nStreamPegBasis{}, the basis from which its drop is measured. \nStreamExtraction{} had been borrowed against the token before the fall, \nStreamExtractionPlume{} of it in one market on Plume; the principal shortfall of all abandoned positions is \nStreamDepositor{}: \nStreamSummaryDepositorUsd{} on replayed Morpho positions, \nStreamStateDepositorUsd{} in Euler vaults and Silo markets on Plasma, mainnet, Arbitrum, Avalanche and Sonic, the largest a Plasma vault that lent USDT0 against xUSD, and \nStreamPtDepositorUsd{} on markets taking its principal token. Lenders of xUSD itself lost \nStreamLoanSide{}, the drop on every unit was \nStreamHolder{}, \nStreamFreeFloatLoss{} of it on the free float that \cref{tab:ledger} charges to holders, and lenders withdrew \nStreamFlight{}. The Plume market now books \nStreamBookShortfall{} of debt, of which \nStreamZombie{} is interest at the pinned rate. A study of this cascade reports \nStreamCompetitorOnchain{} of public lending exposure~\cite{wen2026contagion}: the net stablecoin borrowing of the Morpho and Euler markets that held the failed tokens as bare collateral, one sdeUSD market. On that scope our data agree: the market's book read \nCompetitorSdeusdBook{} at their pre-crisis date, the gap being accrued interest. The rest sits outside that scope: a mainnet Euler vault lending USDC against xUSD and its principal token held \nCompetitorEulerXusd{} at the same date, the loan-side markets above, and the markets on Plume and Plasma that carry most of the shortfall. Public cross-chain estimates reach \nStreamPublicEstimate{}~\cite{blockeden2025stream}; our surface, enumerated from market-creation events on every covered chain, spans \nStreamSurfaceChains{} chains.

\noindent\textit{When the loans were taken.} Over the window from the earliest snapshot, \nStreamWindowBorrow{} was borrowed against the family collateral, \nStreamBorrowBeforeOnsetPct{} of it before the price onset, \nStreamOnset{}, and all but \nStreamBorrowAfterCrossingPct{} before the market price crossed the funded threshold at \nStreamCrossingUtc{} UTC; the largest hour came \nStreamPeakBorrowLeadHours{} hours before the price moved (\cref{fig:timing}d). The Plume market's \nStreamExtractionPlume{} was drawn by one protocol-related address whose pledged units were minted rather than bought (public reporting ties the issuer to borrowing against its own xUSD~\cite{blockeden2025stream}), and \nStreamBorrowAfterPlumeSignalPct{} of the window's volume came after the census's first signal there, a market pinned a week before the price moved.

\subsection{msUSD}\label{app:msusd}

The provider attesting to the reserves ended its coverage~\cite{protos2026msusd} and the price fell to \nMsusdMinPrice{}. On the collateral channel \nMsusdExtraction{} was borrowed; the principal shortfall is \nMsusdDepositor{}, the drop on every unit was \nMsusdHolder{}, \nMsusdFreeFloatLoss{} of it on the free float that \cref{tab:ledger} charges to holders, and \acp{amm} lost \nMsusdLp{}. Lenders withdrew \nMsusdFlight{}: the run, not the outflow, carries this event (\cref{sec:fingerprints}). The damaged market stopped paying its lenders before the coin moved: its lendable balance reached \nMsusdLiquidityAtOnset{} at \nMsusdLiquidityOnsetUtc{}, \nMsusdLiquidityOnsetLeadH{} hours before the price onset, and it stayed at full utilization from then on. The run did not stop at msUSD. The vault that funded the damaged market also funded a mainnet AVLT market and drained it to pay redemptions, and the manager's other vaults, with no exposure at all, were run in the same days~\cite{todayindefi2026alphaping}, one of them lending only against unrelated tokenised credit until its market too sat at full utilization with no bad debt (\cref{sec:fingerprints}). AVLT, the share token of Altura's vault, then lost its own peg: holders rushed to redeem on both chains, the issuer shut the vault~\cite{cryptonews2026altura}, AVLT traded down to \nAvltMinPrice{} while its share value stayed above par, and its market on another chain, which held no msUSD, was drained and pinned. This is the second-order depeg of \cref{sec:vaults}, and AVLT's own market then ran the course of \cref{sec:markets}: loans were taken against the depegged share token, and the market was pinned with \nTfourAvltStock{} of borrowing at the capped rate (\cref{tab:frozen}).

\subsection{USD0++}\label{app:usd0pp}

The issuer set the early-redemption value below par~\cite{usual2025dualexit} and the price fell to \nUsdZeroPPMinPrice{}, \nUsdZeroPPCrossingDepthPct{} below the highest funded liquidation threshold of \nUsdZeroPPFundedLltv{}, and stayed below it for \nUsdZeroPPHoursBelowLltv{} hours: a crossing, but a shallow one. The threshold of \cref{sec:markets} weighs its margin against the entry cost $\kappa$ of buying and pledging a unit, which we estimate from chain state at the trough block, not by fitting: the round trip through the \acp{amm} that trade USD0++ costs \nKappaTroughHundredK{} of the trade at \$100k, \nKappaTroughOneM{} at \$1M, \nKappaTroughTwoM{} at \$2M and \nKappaTroughFiveM{} at \$5M, plus \nKappaGasUsd{} of gas, and crosses the \nUsdZeroPPCrossingDepthPct{} margin near \nKappaBreakevenSize{} of size. A small trade cleared under a cent per dollar and a large one cleared nothing; \nUsdZeroPPExtraction{} was borrowed and abandoned. The drop on every unit was \nUsdZeroPPHolder{}, \nUsdZeroPPFreeFloatLoss{} of it on the free float that \cref{tab:ledger} charges to holders. Lenders withdrew \nUsdZeroPPFlight{}, the largest flight of the five. This is the near-null case for the depositor channel: the price crossed, and the margin it opened was smaller than the cost of using it at any size that would matter.

\section{The Census in Detail}\label{app:census}

\noindent\textit{Rules.} The census covers the markets of the \nCensusProtocols{} protocols (Morpho, Aave v3, Spark, Euler v2, Compound v3, Silo v2, Gearbox and Fluid): \nCensusUniverse{} markets with a pegged token on either side, enumerated from market-creation events rather than token names, which is what catches derived forms. In the USR event \nUsrSurfaceWrappedPct{} of the affected markets list a staked, wrapped or principal-token form and not USR itself. A market is a hit when its oracle price divided by the market price, read at the coin's own \acp{amm} at the blocks the oracle was read, stays at or above $1.08$ for 24 hours together with a sign of stress: utilization at or above 99\% for 72 hours, a borrow rate at or above half the cap of its rate model, or no liquidations. Hits on one token family, a coin with its staked, wrapped and principal-token forms, within \nCensusMergeDays{} days of each other merge into one event across chains. A hit counts as a depeg event only if the coin has a depeg window in the price census (\cref{app:pricecensus}) from \nRnineteenLookbackDays{} days before the hit to its end, or, where the price census cannot price it, a reading at or below $0.97$ on pools holding at least \nRnineteenFloorUsd{} for six hours, and the divergence held in that span. Over \nCensusWindowStart{} to \nCensusWindowEnd{} this leaves \nCensusHits{} hits in \nCensusEvents{} events. \nCensusFalsePositives{} candidate events in which the two prices separated through a unit artifact while the collateral never fell, and \nPinnedWithoutDepegWindows{} market windows pinned while the coin held its peg, are not events. The rules recover all \nSelfcheckCases{} events known in advance.

\noindent\textit{Measurement.} An event is dated from the first six hours in which the market price stays at least $3\%$ below par; the snapshot is the block 24 hours before the first signal on each chain. From it every Morpho and Aave market is replayed per account, exact supply and borrow shares on Morpho and, on Aave, the accounts that supplied the family token and what they borrowed across all reserves. The five cases' \nStateMarkets{} Euler, Silo, Compound and Gearbox markets on the family coin, and the \nPtMarkets{} markets on \nPtChains{} chains that take or lend one of its \nPtTokens{} Pendle principal tokens, are read from state at the snapshot, the onset and the valuation block. An account counts when every collateral it funded is a family token; its principal is its debt less the interest accrued since the snapshot; the depositors' loss is that principal less the collateral at the recovery price, and the outflow of \cref{eq:extract} is the net event-window borrowing by addresses still owing at the end. Coin lenders are the vaults and silos that lend the coin and are not already on the loan side, less the shares the borrowers themselves pledged; \nDualRoleVaults{} Euler vaults that both lend a case coin and take it as collateral are counted on the collateral channel. The recovery price is read at the valuation block under the validity rule of the price census, which \nRecoveryLadderPassed{} of the \nRecoveryTokens{} case tokens pass; where none does, the issuer's redemption or announced settlement (USR and RLP), then a staked form's conversion rate times its underlying, then the median of pools that agree within ten percent, then the largest pool. Five events are rebuilt in full and the rest are sized from market state at their onset. Flows are valued on the day they moved.

\subsection{The price census}\label{app:pricecensus}

The lending census finds a depeg only where a lending market holds the coin. A second scan reads every pegged token of size on its own \acp{amm} over the same window, \nPcWindowFrom{} to \nPcWindowTo{}, whether or not a lending market holds it. Candidates come from two public lists, \nPcSourceLlama{} stablecoins from one and \nPcSourceGecko{} rows over six categories (stablecoins, yield-bearing and bridged stablecoins, liquid staking, liquid restaking and wrapped tokens) from the other, merged by contract address into \nPcUniverse{} tokens, so that two coins of one symbol stay apart and are told by issuer, while a wrapped or staked form is listed under its underlying unless its own pools set the low; \nPcInScope{} remain after dropping tokens whose peak size is below \nPcThreshold{}, tokens with no deployment on a chain of the registry, and tokens whose peg unit has no on-chain numeraire on our chains, and \nPcPriced{} of them have a readable pool. Prices are read on chain, daily and on a \nPcPriceGridHours{}-hour grid around every fall, in the unit of the peg. A reading counts only with money or agreement behind it: the pools behind it hold at least \nPcDepthFloor{} of the sound side at that block, or independent pools agree within ten percent, in which case the reading is the median of the agreeing cluster; the case accounts of \cref{app:events} value their tokens by the same rule. A token is depegged when the price stays below \nPcDepegThreshold{} of par for \nPcSustainHours{} hours, as in the lending census; deployments already below the line when the window opened (\nPcBrokenAtStart{}) are excluded, since their onset lies outside it.

The scan finds \nPcEvents{} depeg windows on \nPcFamilies{} families with a combined peak size of \nPcDepeggedPeakSizeUsd{}. With the families of size that the price rule could not price (the italic entries of \cref{tab:pricecensus}), \nPcUnionSizedFamilies{} families of size depegged, the count of \cref{sec:events}. \Cref{tab:pricecensus} lists them with the holder loss at the lowest price, \nPcHolderPeakUsd{} in all, the loss still standing at the valuation block where a pool of size quotes a price, \nPcHolderRealizedUsd{}, and the \ac{amm} exposure at the onset, \nPcAmmUsd{}. \nPcBucketA{} of the families are events of the lending census, \nPcBucketB{} sat in a census lending market without a lending event, and \nPcBucketC{} were held by no census lending market; the families with a lending event carry \nPcLendingShareOfPeakPct{} of the peak holder loss. The table accounts for each of the \nPcTableLendingEvents{} lending events exactly once.

\input{table7_rows}
\begin{center}
\scriptsize
\setlength{\tabcolsep}{1.6pt}
\begin{longtable}{@{}>{\raggedright\arraybackslash}p{2.1cm} l >{\raggedright\arraybackslash}p{1.7cm} >{\raggedleft\arraybackslash}p{1.5cm} r r >{\raggedleft\arraybackslash}p{1.3cm} r@{}}
\caption{Every coin of size that depegged in the window, by holders' loss at the low, and every lending-census event. In lending: the month of the coin's first lending-census event (+$n$: further events), or \emph{held} where a census lending market held the coin without an event. Lowest price: the lowest valid on-chain price in peg units, or the reason there is none (\emph{low liquidity}: no reading passed the depth or agreement test; \emph{below par at start}: already below the line when its series begins); $\dagger$: priced but outside the sized universe. Supply at onset: units on the deployments that broke. Loss at low, loss now: the holders' loss, supply times the gap from par to the low and to the valuation-block price. Lending: the lending-side loss of \cref{tab:ledger} or \cref{app:census}; \emph{in X}: counted in coin X's row of \cref{tab:ledger}. --: not available or not measured. Last row: totals.}
\label{tab:pricecensus}\\
\toprule
\textbf{Coin} & \textbf{Peg} & \textbf{In lending} & \textbf{Lowest} & \textbf{Supply} & \textbf{Loss at low} & \textbf{Loss now} & \textbf{Lending} \\
\midrule
\endfirsthead
\toprule
\textbf{Coin} & \textbf{Peg} & \textbf{In lending} & \textbf{Lowest} & \textbf{Supply} & \textbf{Loss at low} & \textbf{Loss now} & \textbf{Lending} \\
\midrule
\endhead
\bottomrule
\endfoot
\PriceCensusRows
\end{longtable}
\end{center}

\subsection{The total bound in the five events}\label{app:totalbound}

\Cref{tab:totalbound} gives every term of \cref{eq:eone} for the five events, read at each case's pre-event snapshot: the holders' term $H_c$ (the supply at par over every chain the coin is issued on, bridged copies once, plus the claims denominated in the coin), the credit term ($E^{\mathrm{cred}}_1$, or $E_2$ for the two mints), the liquidation penalties on the debt pledging the coin, the sound side of the \acp{amm} holding it, and the second-order coin (AVLT for msUSD). The \emph{Loss} column is the de-duplicated total of \cref{tab:ledger}. Penalties take each protocol's configured liquidation bonus or discount, the maximum where it varies; the \ac{amm} term takes every census pool of the coin without the depth floor.

\begin{table}[H]
\caption{The total bound of \cref{eq:eone} against the de-duplicated loss, in USD at the valuation date; every term read at the pre-event snapshot.}
\label{tab:totalbound}
\centering
\scriptsize
\setlength{\tabcolsep}{3pt}
\begin{tabular}{@{}l r r r r r r r l@{}}
\toprule
\textbf{Event} & \textbf{Holders $H_c$} & \textbf{Credit} & \textbf{Penalties} & \textbf{\acsp{amm}} & \textbf{Second order} & \textbf{Bound} & \textbf{Loss} & \textbf{Holds} \\
\midrule
rsETH & \nRsethTotalBoundHolders{} & \nRsethTotalBoundCredit{} & \nRsethTotalBoundPenalty{} & \nRsethTotalBoundAmm{} & \nRsethTotalBoundSecondOrder{} & \nRsethTotalBound{} & \nRsethTotalBoundRealized{} & \nRsethTotalBoundHolds{} \\
USR & \nUsrTotalBoundHolders{} & \nUsrTotalBoundCredit{} & \nUsrTotalBoundPenalty{} & \nUsrTotalBoundAmm{} & \nUsrTotalBoundSecondOrder{} & \nUsrTotalBound{} & \nUsrTotalBoundRealized{} & \nUsrTotalBoundHolds{} \\
xUSD & \nStreamTotalBoundHolders{} & \nStreamTotalBoundCredit{} & \nStreamTotalBoundPenalty{} & \nStreamTotalBoundAmm{} & \nStreamTotalBoundSecondOrder{} & \nStreamTotalBound{} & \nStreamTotalBoundRealized{} & \nStreamTotalBoundHolds{} \\
msUSD & \nMsusdTotalBoundHolders{} & \nMsusdTotalBoundCredit{} & \nMsusdTotalBoundPenalty{} & \nMsusdTotalBoundAmm{} & \nMsusdTotalBoundSecondOrder{} & \nMsusdTotalBound{} & \nMsusdTotalBoundRealized{} & \nMsusdTotalBoundHolds{} \\
USD0++ & \nUsdZeroPPTotalBoundHolders{} & \nUsdZeroPPTotalBoundCredit{} & \nUsdZeroPPTotalBoundPenalty{} & \nUsdZeroPPTotalBoundAmm{} & \nUsdZeroPPTotalBoundSecondOrder{} & \nUsdZeroPPTotalBound{} & \nUsdZeroPPTotalBoundRealized{} & \nUsdZeroPPTotalBoundHolds{} \\
\bottomrule
\end{tabular}
\end{table}

\section{Current Exposure across Protocols}\label{app:fulltable}

\Cref{tab:live} gives the bound of \cref{eq:eone} for the largest of the \nLiveBoundCoins{} pegged coins that a covered lending market lists at \nLiveBoundAsOfDate{}, over \nLiveBoundProtocols{} protocols and \nLiveBoundChains{} chains, at four prices. Every term is read from market and reserve totals at that block, as an upper bound: the holders' term from the supply on every chain with bridged copies once, plus the claims in the coin; the lenders' term from each market's idle and routable loan liquidity under its cap, live below par only where the oracle does not move with the coin's own price (a fixed price, an exchange rate, or a feed for another asset); the borrowers' term from each protocol's liquidation penalty on the most debt the coin's collateral can carry; the \acp{amm}' term from every census pool. Markets at full utilization with their rate pinned carry phantom debt and are left out (\nLiveDeadMarkets{}, \nLiveDeadMarketsDebt{} of booked debt). Each coin is its own scenario, so the columns do not add across coins.

\begin{table}[tbp]
\caption{The bound of \cref{eq:eone} for today's pegged coins, in USD as of \nLiveBoundAsOfDate{}: the holders' term $H_c$, the bound at four prices, and the coin's own lenders' term at half of par and at zero; the thirty largest coins by the bound at $p=0.5$, then the rest. At half of par the lenders' term counts only markets whose oracle does not follow the coin's own price (a fixed price, an exchange rate or a feed for another asset), since a market that tracks it liquidates instead; at zero a liquidation recovers nothing and every market counts. Each row is that coin's own scenario, the other coins sound.}
\label{tab:live}
\scriptsize
\setlength{\tabcolsep}{4.5pt}
\begin{tabular}{@{}l r r r r r r r@{}}
\toprule
\textbf{Coin} & $H_c$ & $E(0.9)$ & $E(0.5)$ & $E(0.1)$ & $E(0)$ & \textbf{Lenders (0.5)} & \textbf{Lenders (0)} \\
\midrule
\LiveBoundRows
\bottomrule
\end{tabular}
\end{table}

\end{document}

%% file: numbers.tex
\newcommand{\nCensusEvents}{15}
\newcommand{\nDepositorSide}{\$450.6M}
\newcommand{\nCreditChannel}{\$451.6M}
\newcommand{\nFlightTotal}{\$495.1M}
\newcommand{\nPhantomInterestThree}{\$1.21B}
\newcommand{\nDepositorSideFive}{\$450.6M}
\newcommand{\nUsrExtraction}{\$9.07M}
\newcommand{\nUsrFlight}{\$1.36M}
\newcommand{\nUsrDepositor}{\$8.55M}
\newcommand{\nUsrHolder}{\$108.3M}
\newcommand{\nUsrOnset}{2026-03-22}
\newcommand{\nUsrLoanSide}{\$0.30M}
\newcommand{\nRsethDepositor}{\$195.3M}
\newcommand{\nRsethOnset}{2026-04-19}
\newcommand{\nRsethLoanSide}{\$0}
\newcommand{\nStreamExtraction}{\$33.3M}
\newcommand{\nStreamFlight}{\$58.6M}
\newcommand{\nStreamDepositor}{\$229.3M}
\newcommand{\nStreamHolder}{\$111.1M}
\newcommand{\nStreamZombie}{\$1.05B}
\newcommand{\nStreamOnset}{2025-11-04}
\newcommand{\nStreamLoanSide}{\$0.27M}
\newcommand{\nMsusdExtraction}{\$0.35M}
\newcommand{\nMsusdFlight}{\$24.0M}
\newcommand{\nMsusdDepositor}{\$17.5M}
\newcommand{\nMsusdHolder}{\$39.2M}
\newcommand{\nMsusdOnset}{2026-06-20}
\newcommand{\nMsusdLoanSide}{\$6}
\newcommand{\nUsdZeroPPExtraction}{\$3k}
\newcommand{\nUsdZeroPPFlight}{\$401.8M}
\newcommand{\nUsdZeroPPDepositor}{\$394}
\newcommand{\nUsdZeroPPHolder}{\$38.4M}
\newcommand{\nUsdZeroPPOnset}{2025-01-10}
\newcommand{\nUsdZeroPPLoanSide}{\$3}
\newcommand{\nRecoveryTokens}{10}
\newcommand{\nRecoveryLadderPassed}{4}
\newcommand{\nDualRoleVaults}{14}
\newcommand{\nUsdcGauntletLow}{0.88}
\newcommand{\nUsdcGauntletBadDebt}{\$300k}
\newcommand{\nUsrRecoveryPrice}{0.50}
\newcommand{\nRlpRecoveryPrice}{0.71}
\newcommand{\nUsrSnapLiquidity}{\$0.59M}
\newcommand{\nUsrSnapLtvHeadroom}{\$455.8M}
\newcommand{\nUsrAbandonedDebtOnset}{\$8.15M}
\newcommand{\nUsrAbandonedDebtNow}{\$123.0M}
\newcommand{\nUsrInflowSharePct}{93\%}
\newcommand{\nStreamBookShortfall}{\$1.12B}
\newcommand{\nStreamAbandonedDebtOnset}{\$67.9M}
\newcommand{\nStreamAbandonedDebtNow}{\$1.12B}
\newcommand{\nMsusdAbandonedDebtOnset}{\$24.9M}
\newcommand{\nMsusdAbandonedDebtNow}{\$65.9M}
\newcommand{\nUsdZeroPPFundedLltv}{0.915}
\newcommand{\nInflowUsr}{\$31.1M}
\newcommand{\nRoutableUsr}{\$34.2M}
\newcommand{\nInflowStream}{\$26.3M}
\newcommand{\nRsethAttacker}{\$194.9M}
\newcommand{\nRsethAttackerWeth}{82,620}
\newcommand{\nRsethOfficialWeth}{82,650}
\newcommand{\nRsethNetLoss}{\$195.3M}
\newcommand{\nRsethDeficitWeth}{82,800}
\newcommand{\nRsethLiqRepaid}{\$0}
\newcommand{\nStreamMinPrice}{0.060}
\newcommand{\nMsusdMinPrice}{0.157}
\newcommand{\nUsdZeroPPMinPrice}{0.909}
\newcommand{\nAvltMinPrice}{0.58}
\newcommand{\nFixedPointMarkets}{9}
\newcommand{\nFixedPointEvents}{5}
\newcommand{\nPeakAprPct}{800\%}
\newcommand{\nUsrCappedExtraction}{\$1.20M}
\newcommand{\nUsrCappedHeadroom}{\$14.8M}
\newcommand{\nUsrUncappedExtraction}{\$6.22M}
\newcommand{\nGauntletTotalAssets}{\$118.9M}
\newcommand{\nGauntletMaxLossPct}{21.0\%}
\newcommand{\nGauntletRealizedLossPct}{1.4\%}
\newcommand{\nGauntletCommitted}{\$25.0M}
\newcommand{\nCensusUniverse}{15,081}
\newcommand{\nCensusUniverseChains}{46}
\newcommand{\nCensusProtocols}{8}
\newcommand{\nCensusWindowStart}{2023-08}
\newcommand{\nCensusWindowEnd}{2026-08}
\newcommand{\nCensusFalsePositives}{39}
\newcommand{\nPinnedWithoutDepegWindows}{173}
\newcommand{\nRnineteenLookbackDays}{30}
\newcommand{\nRnineteenFloorUsd}{\$100}
\newcommand{\nCensusHits}{64}
\newcommand{\nCensusMergeDays}{7}
\newcommand{\nSelfcheckCases}{5}
\newcommand{\nUsrCrossingUtc}{2026-03-22 02:35}
\newcommand{\nUsrBorrowAfterCrossingPct}{100\%}
\newcommand{\nUsrPeakBorrowLeadHours}{0}
\newcommand{\nMsusdBorrowAfterCrossingPct}{95\%}
\newcommand{\nUsdZeroPPHoursBelowLltv}{15}
\newcommand{\nStreamCrossingUtc}{2025-11-03 20:00}
\newcommand{\nStreamBorrowAfterCrossingPct}{0.1\%}
\newcommand{\nStreamWindowBorrow}{\$34.3M}
\newcommand{\nStreamPeakBorrowLeadHours}{258}
\newcommand{\nStreamBorrowBeforeOnsetPct}{100.0\%}
\newcommand{\nStreamBorrowAfterPlumeSignalPct}{71.6\%}
\newcommand{\nUsrSurfaceMarkets}{271}
\newcommand{\nUsrSurfaceChains}{6}
\newcommand{\nStreamSurfaceChains}{7}
\newcommand{\nUsrSurfaceWrappedPct}{49\%}
\newcommand{\nUsrExtractionMainnetMorpho}{\$7.74M}
\newcommand{\nUsrExtractionBase}{\$1.25M}
\newcommand{\nStreamExtractionPlume}{\$32.0M}
\newcommand{\nRsethSeized}{89,567}
\newcommand{\nUsrLp}{\$0.30M}
\newcommand{\nMsusdLp}{\$0.76M}
\newcommand{\nUsrFrozenMDays}{5,513.1}
\newcommand{\nStreamFrozenMDays}{142.3}
\newcommand{\nMsusdFrozenMDays}{804.6}
\newcommand{\nUsdZeroPPFrozenMDays}{0.4}
\newcommand{\nFrozenTotalMDays}{6,460.5}
\newcommand{\nFrozenEvents}{14}
\newcommand{\nPinnedEvents}{9}
\newcommand{\nTfourUsrStock}{\$11.6M}
\newcommand{\nTfourUsrHours}{2,399}
\newcommand{\nTfourUsrApr}{800\%}
\newcommand{\nTfourUsrCap}{4}
\newcommand{\nTfourUsrMarkets}{17}
\newcommand{\nTfourUsrMDays}{555.8}
\newcommand{\nTfourWstUsrStock}{\$69.3M}
\newcommand{\nTfourWstUsrHours}{320}
\newcommand{\nTfourWstUsrApr}{800\%}
\newcommand{\nTfourWstUsrCap}{1}
\newcommand{\nTfourWstUsrMarkets}{4}
\newcommand{\nTfourWstUsrMDays}{924.5}
\newcommand{\nTfourRlpStock}{\$29.1M}
\newcommand{\nTfourRlpHours}{3,257}
\newcommand{\nTfourRlpApr}{800\%}
\newcommand{\nTfourRlpCap}{2}
\newcommand{\nTfourRlpMarkets}{5}
\newcommand{\nTfourRlpMDays}{4,032.9}
\newcommand{\nTfourMsusdStock}{\$57.9M}
\newcommand{\nTfourMsusdHours}{265}
\newcommand{\nTfourMsusdApr}{800\%}
\newcommand{\nTfourMsusdCap}{1}
\newcommand{\nTfourMsusdMarkets}{2}
\newcommand{\nTfourMsusdMDays}{682.9}
\newcommand{\nTfourAvltStock}{\$2.87M}
\newcommand{\nTfourAvltHours}{100}
\newcommand{\nTfourAvltApr}{800\%}
\newcommand{\nTfourAvltCap}{1}
\newcommand{\nTfourAvltMarkets}{1}
\newcommand{\nTfourAvltMDays}{121.8}
\newcommand{\nTfourXusdStock}{\$14.4M}
\newcommand{\nTfourXusdHours}{238}
\newcommand{\nTfourXusdApr}{56\%}
\newcommand{\nTfourXusdCap}{0}
\newcommand{\nTfourXusdMarkets}{1}
\newcommand{\nTfourXusdMDays}{142.3}
\newcommand{\nTfourUsdZeroPPStock}{\$54k}
\newcommand{\nTfourUsdZeroPPHours}{12,406}
\newcommand{\nTfourUsdZeroPPApr}{225\%}
\newcommand{\nTfourUsdZeroPPCap}{0}
\newcommand{\nTfourUsdZeroPPMarkets}{3}
\newcommand{\nTfourUsdZeroPPMDays}{0.4}
\newcommand{\nTfourRsethStock}{\$0}
\newcommand{\nTfourRsethHours}{0}
\newcommand{\nTfourRsethApr}{29\%}
\newcommand{\nTfourRsethCap}{0}
\newcommand{\nTfourRsethMarkets}{1}
\newcommand{\nTfourRsethMDays}{0.0}
\newcommand{\nStreamPegBasis}{1.266}
\newcommand{\nNovelOtherLoss}{\$0.35M}
\newcommand{\nNovelOtherDepositorUsd}{\$5k}
\newcommand{\nNovelOtherHolderUsd}{\$63.0M}
\newcommand{\nNovelOtherAmmUsd}{\$5.31M}
\newcommand{\nNovelOtherDedupUsd}{\$68.7M}
\newcommand{\nBoundSelUsr}{\$34.6M}
\newcommand{\nBoundSelRseth}{\$1.37B}
\newcommand{\nBoundSelUseRsethPct}{16\%}
\newcommand{\nBoundSelStream}{\$3.76M}
\newcommand{\nUsrTotalBound}{\$195.9M}
\newcommand{\nUsrTotalBoundHolders}{\$146.5M}
\newcommand{\nUsrTotalBoundCredit}{\$34.6M}
\newcommand{\nUsrTotalBoundPenalty}{\$1.79M}
\newcommand{\nUsrTotalBoundAmm}{\$13.1M}
\newcommand{\nUsrTotalBoundSecondOrder}{\$0}
\newcommand{\nUsrTotalBoundRealized}{\$88.5M}
\newcommand{\nUsrTotalBoundHolds}{yes}
\newcommand{\nRsethTotalBound}{\$2.94B}
\newcommand{\nRsethTotalBoundHolders}{\$1.55B}
\newcommand{\nRsethTotalBoundCredit}{\$1.37B}
\newcommand{\nRsethTotalBoundPenalty}{\$15.0M}
\newcommand{\nRsethTotalBoundAmm}{\$2.15M}
\newcommand{\nRsethTotalBoundSecondOrder}{\$0}
\newcommand{\nRsethTotalBoundRealized}{\$195.3M}
\newcommand{\nRsethTotalBoundHolds}{yes}
\newcommand{\nStreamTotalBound}{\$381.7M}
\newcommand{\nStreamTotalBoundHolders}{\$346.4M}
\newcommand{\nStreamTotalBoundCredit}{\$3.76M}
\newcommand{\nStreamTotalBoundPenalty}{\$31.0M}
\newcommand{\nStreamTotalBoundAmm}{\$0.55M}
\newcommand{\nStreamTotalBoundSecondOrder}{\$0}
\newcommand{\nStreamTotalBoundRealized}{\$256.4M}
\newcommand{\nStreamTotalBoundHolds}{yes}
\newcommand{\nMsusdTotalBound}{\$127.1M}
\newcommand{\nMsusdTotalBoundHolders}{\$82.2M}
\newcommand{\nMsusdTotalBoundCredit}{\$3.75M}
\newcommand{\nMsusdTotalBoundPenalty}{\$1.67M}
\newcommand{\nMsusdTotalBoundAmm}{\$0.40M}
\newcommand{\nMsusdTotalBoundSecondOrder}{\$39.1M}
\newcommand{\nMsusdTotalBoundRealized}{\$39.5M}
\newcommand{\nMsusdTotalBoundHolds}{yes}
\newcommand{\nUsdZeroPPTotalBound}{\$1.72B}
\newcommand{\nUsdZeroPPTotalBoundHolders}{\$1.51B}
\newcommand{\nUsdZeroPPTotalBoundCredit}{\$40.1M}
\newcommand{\nUsdZeroPPTotalBoundPenalty}{\$15.2M}
\newcommand{\nUsdZeroPPTotalBoundAmm}{\$148.9M}
\newcommand{\nUsdZeroPPTotalBoundSecondOrder}{\$0}
\newcommand{\nUsdZeroPPTotalBoundRealized}{\$38.4M}
\newcommand{\nUsdZeroPPTotalBoundHolds}{yes}
\newcommand{\nLiveBoundCoins}{137}
\newcommand{\nLiveBoundProtocols}{7}
\newcommand{\nLiveBoundChains}{39}
\newcommand{\nLiveBoundAsOfDate}{2026-08-30}
\newcommand{\nLiveDeadMarkets}{56}
\newcommand{\nLiveDeadMarketsDebt}{\$4.97B}
\newcommand{\nLiveBoundLenderTopOneCoin}{wstETH}
\newcommand{\nLiveBoundLenderTopOneValue}{\$1.05B}
\newcommand{\nLiveBoundLenderTopTwoCoin}{weETH}
\newcommand{\nLiveBoundLenderTopTwoValue}{\$809.7M}
\newcommand{\nLiveBoundLenderTopThreeCoin}{WETH}
\newcommand{\nLiveBoundLenderTopThreeValue}{\$678.3M}
\newcommand{\nLiveBoundLenderZeroTopOneCoin}{wstETH}
\newcommand{\nLiveBoundLenderZeroTopOneValue}{\$2.80B}
\newcommand{\nLiveBoundRowsListed}{30}
\newcommand{\nLiveBoundTopOneCoin}{USDT}
\newcommand{\nLiveBoundTopOneAtFifty}{\$52.83B}
\newcommand{\nLiveBoundRankListedLastCoin}{mETH}
\newcommand{\nLiveBoundRankListedLastAtFifty}{\$291.2M}
\newcommand{\nStreamCompetitorOnchain}{\$7.60M}
\newcommand{\nStreamPublicEstimate}{\$285.0M}
\newcommand{\nCompetitorSdeusdBook}{\$8.65M}
\newcommand{\nCompetitorEulerXusd}{\$14.3M}
\newcommand{\nValuationDate}{2026-09-10}
\newcommand{\nUncoveredAfterInflowStream}{\$3.20M}
\newcommand{\nDedupTotal}{\$686.8M}
\newcommand{\nDedupTotalFive}{\$618.1M}
\newcommand{\nDedupFreeFloat}{\$188.7M}
\newcommand{\nDedupBorrower}{\$33.5M}
\newcommand{\nDedupBorrowerEquity}{\$3.47M}
\newcommand{\nUsrDedup}{\$88.5M}
\newcommand{\nUsrFreeFloatLoss}{\$58.2M}
\newcommand{\nUsrDedupLp}{\$0.30M}
\newcommand{\nUsrBorrowerLoss}{\$21.1M}
\newcommand{\nRsethDedup}{\$195.3M}
\newcommand{\nRsethFreeFloatLoss}{\$0}
\newcommand{\nRsethDedupLp}{\$0}
\newcommand{\nRsethBorrowerLoss}{\$0}
\newcommand{\nStreamDedup}{\$256.4M}
\newcommand{\nStreamFreeFloatLoss}{\$25.6M}
\newcommand{\nStreamDedupLp}{\$1.07M}
\newcommand{\nStreamBorrowerLoss}{\$0.15M}
\newcommand{\nMsusdDedup}{\$39.5M}
\newcommand{\nMsusdFreeFloatLoss}{\$18.4M}
\newcommand{\nMsusdDedupLp}{\$0.76M}
\newcommand{\nMsusdBorrowerLoss}{\$2.93M}
\newcommand{\nUsdZeroPPDedup}{\$38.4M}
\newcommand{\nUsdZeroPPFreeFloatLoss}{\$23.5M}
\newcommand{\nUsdZeroPPDedupLp}{\$5.67M}
\newcommand{\nUsdZeroPPBorrowerLoss}{\$9.23M}
\newcommand{\nKappaTroughHundredK}{0.06\%}
\newcommand{\nKappaTroughOneM}{0.51\%}
\newcommand{\nKappaTroughTwoM}{0.96\%}
\newcommand{\nKappaTroughFiveM}{2.11\%}
\newcommand{\nKappaGasUsd}{\$5}
\newcommand{\nUsdZeroPPCrossingDepthPct}{0.66\%}
\newcommand{\nKappaBreakevenSize}{\$1.33M}
\newcommand{\nFigFiveCuratorTroughPct}{31\%}
\newcommand{\nFigFiveAvltPinnedHours}{24}
\newcommand{\nFigFiveCoVaultedTroughPct}{63\%}
\newcommand{\nFigFiveCoVaultedPinnedHours}{48}
\newcommand{\nFigFiveManagerCleanPinnedHours}{48}
\newcommand{\nFigFiveCoAfterPinGrowthPct}{16\%}
\newcommand{\nFigFiveCoVaultedIdleOnsetPct}{9\%}
\newcommand{\nFigFiveCtrlIdleOnsetPct}{11\%}
\newcommand{\nFigFiveCtrlIdleMinPct}{10\%}
\newcommand{\nFigFiveCuratorChain}{HyperEVM}
\newcommand{\nPartRsethEmbPct}{88.8\%}
\newcommand{\nPartRsethLendPct}{88.6\%}
\newcommand{\nPartRsethDexPct}{0.2\%}
\newcommand{\nPartRlpEmbPct}{55.4\%}
\newcommand{\nPartRlpLendPct}{55.1\%}
\newcommand{\nPartRlpDexPct}{0.3\%}
\newcommand{\nPartMsusdEmbPct}{53.2\%}
\newcommand{\nPartMsusdLendPct}{49.2\%}
\newcommand{\nPartMsusdDexPct}{4.0\%}
\newcommand{\nPartUsdZeroPPEmbPct}{38.8\%}
\newcommand{\nPartUsdZeroPPLendPct}{24.0\%}
\newcommand{\nPartUsdZeroPPDexPct}{14.8\%}
\newcommand{\nPartXusdEmbPct}{16.6\%}
\newcommand{\nPartXusdLendPct}{15.6\%}
\newcommand{\nPartXusdDexPct}{1.0\%}
\newcommand{\nPartUsrEmbPct}{17.9\%}
\newcommand{\nPartUsrLendPct}{17.3\%}
\newcommand{\nPartUsrDexPct}{0.6\%}
\newcommand{\nPartRsethAavePct}{83.5\%}
\newcommand{\nPartUsrBridgePct}{51.5\%}
\newcommand{\nDedupPools}{\$13.1M}
\newcommand{\nDedupCreditPct}{66\%}
\newcommand{\nDedupFreeFloatPct}{27\%}
\newcommand{\nDedupBorrowerPct}{5\%}
\newcommand{\nDedupPoolsPct}{2\%}
\newcommand{\nDedupSelfCreatedUsd}{\$136.6M}
\newcommand{\nUsrSelfCreatedHolderUsd}{\$24.2M}
\newcommand{\nRsethForgedUnits}{116,500}
\newcommand{\nUsrPtDepositorUsd}{\$0.42M}
\newcommand{\nUsrStateDepositorUsd}{\$0.46M}
\newcommand{\nUsrSummaryDepositorUsd}{\$7.66M}
\newcommand{\nStreamPtDepositorUsd}{\$0.38M}
\newcommand{\nStreamStateDepositorUsd}{\$161.1M}
\newcommand{\nStreamSummaryDepositorUsd}{\$67.8M}
\newcommand{\nPtTokens}{75}
\newcommand{\nPtChains}{9}
\newcommand{\nPtMarkets}{120}
\newcommand{\nStateMarkets}{245}
\newcommand{\nLoanSideAll}{\$0.93M}
\newcommand{\nRsethAipDate}{2026-05-06}
\newcommand{\nRsethAttackerWsteth}{821}
\newcommand{\nRsethOfficialWsteth}{821}
\newcommand{\nRsethDeficitWsteth}{821}
\newcommand{\nRsethEthOnsetPrice}{2,330}
\newcommand{\nMsusdLiquidityOnsetUtc}{2026-06-17 18:00}
\newcommand{\nMsusdLiquidityOnsetLeadH}{70}
\newcommand{\nMsusdLiquidityAtOnset}{\$0}
\newcommand{\nMsusdPreOnsetWithdrawn}{\$7.27M}
\newcommand{\nCuratorRunFamilyDrawdownPct}{65.0\%}
\newcommand{\nCensusPriceGridHours}{4}
\newcommand{\nPcPriceGridHours}{6}
\newcommand{\nPcWindowFrom}{2023-08}
\newcommand{\nPcWindowTo}{2026-08}
\newcommand{\nPcWindowFromMonth}{August 2023}
\newcommand{\nPcWindowToMonth}{August 2026}
\newcommand{\nPcThreshold}{\$10.0M}
\newcommand{\nPcDepegThreshold}{0.97}
\newcommand{\nPcSustainHours}{6}
\newcommand{\nPcSourceLlama}{427}
\newcommand{\nPcSourceGecko}{1,228}
\newcommand{\nPcUniverse}{1,402}
\newcommand{\nPcInScope}{347}
\newcommand{\nPcPriced}{295}
\newcommand{\nPcEvents}{56}
\newcommand{\nPcFamilies}{30}
\newcommand{\nPcBucketA}{5}
\newcommand{\nPcBucketB}{13}
\newcommand{\nPcBucketC}{12}
\newcommand{\nPcDepeggedPeakSizeUsd}{\$7.94B}
\newcommand{\nPcHolderPeakUsd}{\$614.0M}
\newcommand{\nPcHolderRealizedUsd}{\$104.0M}
\newcommand{\nPcAmmUsd}{\$49.5M}
\newcommand{\nPcLendingShareOfPeakPct}{63.4\%}
\newcommand{\nPcUnionPeakSizeUsd}{\$10.15B}
\newcommand{\nPcUnionShareOfSizePct}{2.1\%}
\newcommand{\nPcUnionSizedFamilies}{34}
\newcommand{\nPcBrokenAtStart}{22}
\newcommand{\nPcDepthFloor}{\$10k}
\newcommand{\nPcTableLendingEvents}{15}
\newcommand{\nCapMarketsBound}{14}
\newcommand{\nCapMarketsWithOutflow}{71}
\newcommand{\nCapRsethMainnetRoomPct}{6.1\%}
\newcommand{\nCapRsethMainnetFilledUtc}{2026-04-18 17:50 UTC}

%% file: table_live_rows.tex
\newcommand{\LiveBoundRows}{%
USDT & \$105.05B & \$10.68B & \$52.83B & \$94.98B & \$107.54B & \$53.8M & \$2.13B \\
USDC & \$76.49B & \$7.81B & \$38.56B & \$69.32B & \$78.96B & \$84.0M & \$2.13B \\
stETH & \$23.67B & \$3.93B & \$19.10B & \$34.63B & \$38.53B & \$0 & \$2.05M \\
wstETH & \$11.69B & \$1.57B & \$7.26B & \$13.33B & \$14.86B & \$1.05B & \$2.80B \\
WETH & \$11.02B & \$1.39B & \$6.48B & \$11.74B & \$13.06B & \$678.3M & \$1.74B \\
cbBTC & \$8.34B & \$1.03B & \$4.98B & \$9.28B & \$10.62B & \$609.9M & \$2.07B \\
WBTC & \$9.19B & \$1.10B & \$4.80B & \$8.51B & \$11.50B & \$17.8M & \$2.11B \\
USDS & \$7.61B & \$767.7M & \$3.83B & \$6.90B & \$7.67B & \$10.1M & \$29.4M \\
weETH & \$4.95B & \$737.5M & \$3.52B & \$6.69B & \$7.51B & \$809.7M & \$2.31B \\
USDe & \$4.73B & \$596.6M & \$2.77B & \$4.95B & \$5.55B & \$262.1M & \$637.6M \\
DAI & \$5.00B & \$516.0M & \$2.56B & \$4.61B & \$5.12B & \$0 & \$11k \\
PYUSD & \$2.68B & \$321.0M & \$1.59B & \$2.86B & \$3.19B & \$28.7M & \$69.0M \\
USD1 & \$2.96B & \$297.8M & \$1.48B & \$2.67B & \$2.97B & \$0 & \$0 \\
USYC & \$2.78B & \$277.6M & \$1.39B & \$2.50B & \$2.78B & \$2.27M & \$5.11M \\
USDG & \$2.29B & \$231.5M & \$1.16B & \$2.08B & \$2.32B & \$0 & \$9k \\
crvUSD & \$2.10B & \$213.8M & \$1.07B & \$1.93B & \$2.14B & \$0 & \$0 \\
rsETH & \$1.10B & \$202.4M & \$1.04B & \$1.88B & \$2.09B & \$397.9M & \$862.0M \\
RLUSD & \$1.73B & \$189.4M & \$911.9M & \$1.63B & \$1.82B & \$0 & \$7.73M \\
USDY & \$1.42B & \$142.1M & \$710.7M & \$1.28B & \$1.42B & \$0 & \$0 \\
USDf & \$1.29B & \$129.5M & \$647.4M & \$1.17B & \$1.29B & \$23k & \$52k \\
cbETH & \$1.10B & \$124.4M & \$600.7M & \$1.09B & \$1.41B & \$35.3M & \$288.6M \\
GHO & \$1.05B & \$109.8M & \$561.1M & \$1.02B & \$1.13B & \$31.8M & \$76.5M \\
rETH & \$921.2M & \$104.0M & \$536.9M & \$970.7M & \$1.08B & \$68.8M & \$150.5M \\
LBTC & \$677.4M & \$100.3M & \$525.4M & \$1.01B & \$1.13B & \$136.7M & \$380.6M \\
BUIDL & \$954.7M & \$95.5M & \$477.3M & \$859.2M & \$954.7M & \$1k & \$3k \\
LsETH & \$785.3M & \$78.8M & \$394.6M & \$710.3M & \$789.2M & \$39k & \$93k \\
USDX & \$685.4M & \$70.7M & \$344.9M & \$619.0M & \$687.5M & \$0 & \$20 \\
osETH & \$365.6M & \$62.4M & \$339.2M & \$616.0M & \$685.2M & \$140.4M & \$303.6M \\
kBTC & \$583.0M & \$74.0M & \$332.9M & \$593.9M & \$659.2M & \$18.3M & \$43.8M \\
mETH & \$580.6M & \$58.2M & \$291.2M & \$524.2M & \$582.5M & \$0 & \$197 \\
\midrule
Other 107 coins & \$8.66B & \$962.5M & \$4.71B & \$8.47B & \$9.53B & \$70.0M & \$300.5M \\
}

%% file: table7_rows.tex
\newcommand{\PriceCensusRows}{%
USR & USD & 2026-03, +1 & 0.166 & 183.1M & \$152.7M & -- & \$8.85M \\
apxUSD & USD & 2026-06 & 0.750 & 514.6M & \$128.4M & \$5.12M & \$50k \\
pmUSD & USD & 2026-05 & 0.360 & 100.2M & \$64.1M & \$50.2M & -- \\
sUSD & USD & held & 0.238 & 62.9M & \$47.9M & -- & -- \\
wrsETH & ETH & 2026-04 & 0.505 & 30k & \$35.1M & \$0 & in rsETH \\
YU & USD & -- & 0.719 & 116.5M & \$32.7M & -- & -- \\
USDR & USD & held & 0.016 & 21.1M & \$20.7M & \$20.7M & -- \\
EURT & EUR & -- & 0.680 & 50.0M & \$18.8M & -- & -- \\
USDA & USD & -- & 0.815 & 87.0M & \$16.1M & \$6.09M & -- \\
uniBTC & BTC & held & 0.814 & 1k & \$14.6M & \$0.78M & -- \\
TUSD & USD & held & 0.963 & 390.0M & \$14.6M & \$0.71M & -- \\
USDz & USD & held & 0.889 & 102.5M & \$11.4M & \$3.04M & -- \\
miMATIC & USD & -- & 0.969 & 305.1M & \$9.45M & \$6.16M & -- \\
msUSD (Metronome) & USD & 2026-07 & 0.600 & 22.9M & \$9.15M & \$7.67M & \$0.31M \\
DOLA & USD & held & 0.963 & 204.8M & \$7.58M & \$56k & -- \\
PUSD & USD & held & 0.939 & 120.0M & \$7.37M & -- & -- \\
sAID & USD & -- & 0.796 & 20.9M & \$4.25M & -- & -- \\
rswETH & ETH & held & 0.819 & 12k & \$3.38M & \$3.38M & -- \\
sUSDat & USD & held & 0.963 & 80.7M & \$2.99M & \$0 & -- \\
lisUSD & USD & held & 0.957 & 68.9M & \$2.97M & \$0.12M & -- \\
AID & USD & -- & 0.959 & 72.6M & \$2.96M & -- & -- \\
M-BTC & BTC & held & 0.959 & 500 & \$2.19M & \$0 & -- \\
aUSDT & USD & -- & 0.969 & 50.0M & \$1.53M & -- & -- \\
ULTRA & USD & -- & 0.936 & 22.5M & \$1.43M & \$0 & -- \\
peUSD & USD & -- & 0.939 & 11.3M & \$0.69M & -- & -- \\
XAI & USD & -- & 0.964 & 12.2M & \$0.44M & -- & -- \\
USDU, Ethereum & USD & held & 0.960 & 5.3M & \$0.21M & \$0 & -- \\
brBTC & BTC & -- & 0.934 & 25 & \$0.16M & -- & -- \\
GRAI & USD & -- & 0.961 & 2.9M & \$0.11M & -- & -- \\
eUSD & USD & held & 0.930 & 159k & \$11k & \$0 & -- \\
bUSD0 & USD & 2025-01 & 0.909$^{\dagger}$ & 1.46B & \$132.4M & \$38.8M & \$398 \\
AVLT & USD & 2026-06 & 0.126$^{\dagger}$ & 29.6M & \$25.9M & \$21.5M & in msUSD \\
RLP & USD & 2026-03 & 0.349$^{\dagger}$ & 30.0M & \$19.5M & -- & in USR \\
ynRWAx & USD & 2026-07 & 0.866$^{\dagger}$ & 8.7M & \$1.16M & -- & -- \\
rsETH & ETH & 2026-04, +1 & \textit{low liquidity} & -- & -- & -- & \$195.3M \\
xUSD & USD & 2025-11 & \textit{below par at start} & -- & -- & -- & \$229.6M \\
USDX & USD & 2025-11 & \textit{low liquidity} & -- & -- & -- & -- \\
msUSD (Main Street) & USD & 2026-06 & \textit{low liquidity} & -- & -- & -- & \$17.5M \\
\midrule
All & -- & 15 events & -- & -- & \$614.0M & \$104.0M & -- \\
}